\documentclass[twocolumn]{aastex631}
\usepackage{amsmath,amssymb,amstext}
\usepackage{gensymb}
\usepackage{multirow}
\usepackage{rotating}
\usepackage{tabularx}

\usepackage{xspace}

\graphicspath{{./}{figures/}}

\newcommand{\hst}{{\textit{HST}}\xspace}
\newcommand{\jwst}{{\textit{JWST}}\xspace}

\newcommand{\planetname}{{GJ~1132~b}\xspace}

\newcommand{\planetrad}{{1.30\,R$_{\oplus}$}\xspace}

\newcommand{\planetmass}{{1.66\,M$_{\oplus}$}\xspace}

\newcommand{\planettemp}{{529\,K}\xspace}

\newcommand{\starrad}{{0.2105\,R$_{\odot}$}\xspace}

\newcommand{\startemp}{{3261\,K\xspace}}

\newcommand{\wavestart}{2.8\xspace}
\newcommand{\waveend}{5.2\xspace}

\newcommand{\ngroups}{14\xspace}
\newcommand{\nints}{814\xspace}
\newcommand{\obstime}{3.06\xspace}

\newcommand{\eureka}{\texttt{Eureka!}\xspace}
\newcommand{\firefly}{\texttt{FIREFLy}\xspace}
\newcommand{\exotic}{\texttt{ExoTiC-JEDI}\xspace}

\newcommand{\POSEIDON}{\texttt{POSEIDON}\xspace}

\defcitealias{LustigYaeger2023}{Lustig-Yaeger \& Fu et al.}
\defcitealias{Moran2023}{Moran \& Stevenson et al.}

\makeatletter
\newcommand*{\linktocite}[2]{%
  \hyper@natlinkstart{#1}#2\hyper@natlinkend}
\makeatother

\submitjournal{ApJL}

\shorttitle{GJ 1132 b keeps its secrets}
\shortauthors{May \& MacDonald et al.}

\begin{document}

\title{Double Trouble: Two Transits of the Super-Earth GJ 1132 b Observed with JWST NIRSpec G395H}

\correspondingauthor{Erin M. May \& Ryan J. MacDonald }
\email{Erin.May@jhuapl.edu, ryanjmac@umich.edu}

\author[0000-0002-2739-1465]{E. M. May}
\altaffiliation{These authors contributed equally to this work.}
\affil{Johns Hopkins APL, Laurel, MD 20723, USA}
\affil{Consortium on Habitability and Atmospheres of M-dwarf Planets (CHAMPs), Laurel, MD, USA}

\author[0000-0003-4816-3469]{Ryan J. MacDonald}
\altaffiliation{These authors contributed equally to this work.}
\affil{Department of Astronomy, University of Michigan, Ann Arbor, MI 48109, USA}
\affil{NHFP Sagan Fellow}

\author[0000-0002-9030-0132]{Katherine A. Bennett}
\affiliation{Department of Earth \& Planetary Sciences, Johns Hopkins University, Baltimore, MD, USA}

\author[0000-0002-6721-3284]{Sarah E. Moran}
\affiliation{Department of Planetary Sciences and Lunar and Planetary Laboratory, University of Arizona, Tuscon, AZ, USA}

\author[0000-0003-4328-3867]{Hannah R. Wakeford}
\affil{School of Physics, HH Wills Physics Laboratory, University of Bristol, Bristol, UK}
\affil{Consortium on Habitability and Atmospheres of M-dwarf Planets (CHAMPs), Laurel, MD, USA}

\author[0000-0002-1046-025X]{Sarah Peacock}
\affil{University of Maryland, Baltimore County, MD 21250, USA}
\affil{NASA Goddard Space Flight Center, Greenbelt, MD 20771, USA}
\affil{Consortium on Habitability and Atmospheres of M-dwarf Planets (CHAMPs), Laurel, MD, USA}

\author[0000-0002-0746-1980]{Jacob Lustig-Yaeger}
\affil{Johns Hopkins APL, Laurel, MD 20723, USA}
\affil{Consortium on Habitability and Atmospheres of M-dwarf Planets (CHAMPs), Laurel, MD, USA}

\author[0009-0009-3217-0403]{Alicia N. Highland}
\affil{Department of Astronomy, University of Michigan, Ann Arbor, MI 48109, USA}

\author[0000-0002-7352-7941]{Kevin B. Stevenson}
\affiliation{Johns Hopkins APL, Laurel, MD 20723, USA}
\affil{Consortium on Habitability and Atmospheres of M-dwarf Planets (CHAMPs), Laurel, MD, USA}

\author[0000-0001-6050-7645]{David K. Sing}
\affiliation{Department of Physics \& Astronomy, Johns Hopkins University, Baltimore, MD, USA}
\affiliation{Department of Earth \& Planetary Sciences, Johns Hopkins University, Baltimore, MD, USA}

\author[0000-0002-4321-4581]{L. C. Mayorga}
\affiliation{Johns Hopkins APL, Laurel, MD 20723, USA}
\affil{Consortium on Habitability and Atmospheres of M-dwarf Planets (CHAMPs), Laurel, MD, USA}

\author[0000-0003-1240-6844]{Natasha E. Batalha}
\affiliation{NASA Ames Research Center, Moffett Field, CA 94035, USA}
\affil{Consortium on Habitability and Atmospheres of M-dwarf Planets (CHAMPs), Laurel, MD, USA}

\author[0000-0002-4207-6615]{James Kirk}
\affil{Department of Physics, Imperial College London, Prince Consort Road, London, SW7 2AZ, UK}

\author[0000-0003-3204-8183]{Mercedes L\'opez-Morales}
\affiliation{Center for Astrophysics ${\rm \mid}$ Harvard {\rm \&} Smithsonian, 60 Garden St, Cambridge, MA 02138, USA}
\affil{Consortium on Habitability and Atmospheres of M-dwarf Planets (CHAMPs), Laurel, MD, USA}

\author[0000-0003-3305-6281]{Jeff A. Valenti}
\affiliation{Space Telescope Science Institute, Baltimore, MD 21218, USA}
\affil{Consortium on Habitability and Atmospheres of M-dwarf Planets (CHAMPs), Laurel, MD, USA}

\author[0000-0003-4157-832X]{Munazza K. Alam}
\affil{Carnegie Earth \& Planets Laboratory, Washington, DC, 20015 USA}

\author[0000-0001-8703-7751]{Lili Alderson}
\affiliation{School of Physics, HH Wills Physics Laboratory, University of Bristol, Bristol, UK}

\author[0000-0002-3263-2251]{Guangwei Fu}
\affil{Department of Physics \& Astronomy, Johns Hopkins University, Baltimore, MD, USA}

\author[0000-0002-9032-8530]{Junellie Gonzalez-Quiles}
\affiliation{Department of Earth \& Planetary Sciences, Johns Hopkins University, Baltimore, MD, USA}
\affil{Consortium on Habitability and Atmospheres of M-dwarf Planets (CHAMPs), Laurel, MD, USA}

\author[0000-0003-3667-8633]{Joshua D. Lothringer}
\affil{Department of Physics, Utah Valley University, Orem, UT, 84058 USA}

\author[0000-0003-4408-0463]{Zafar Rustamkulov}
\affiliation{Department of Earth \& Planetary Sciences, Johns Hopkins University, Baltimore, MD, USA}

\author[0000-0001-7393-2368]{Kristin S. Sotzen}
\affiliation{Johns Hopkins APL, Laurel, MD 20723, USA}
\affil{Consortium on Habitability and Atmospheres of M-dwarf Planets (CHAMPs), Laurel, MD, USA}

\begin{abstract}
The search for rocky planet atmospheres with JWST has focused on planets transiting M dwarfs. Such planets have favorable planet-to-star size ratios, enhancing the amplitude of atmospheric features. Since the expected signal strength of atmospheric features is similar to the single-transit performance of \jwst, multiple observations are required to confirm any detection.  Here, we present two transit observations of the rocky planet \planetname with \jwst NIRSpec G395H, covering \wavestart --\waveend\,$\micron$. Previous \hst WFC3 observations of \planetname were inconclusive, with evidence reported for either an atmosphere or a featureless spectrum based on analyses of the same dataset. Our \jwst data exhibit substantial differences between the two visits. One transit is consistent with either a H$_2$O-dominated atmosphere containing $\sim$ 1\% CH$_4$ and trace N$_2$O ($\chi^2_{\nu} = 1.13$) or stellar contamination from unocculted starspots ($\chi^2_{\nu} = 1.36$). However, the second transit is consistent with a featureless spectrum. Neither visit is consistent with a previous report of HCN. Atmospheric variability is unlikely to explain the scale of the observed differences between the visits. Similarly, our out-of-transit stellar spectra show no evidence of changing stellar inhomogeneity between the two visits --- observed 8 days apart, only 6.5\% of the stellar rotation rate. We further find no evidence of differing instrumental systematic effects between visits. The most plausible explanation is an unlucky random noise draw leading to two significantly discrepant transmission spectra. Our results highlight the importance of multi-visit repeatability with \jwst prior to claiming atmospheric detections for these small, enigmatic planets.
\end{abstract}

\section{Introduction} \label{sec:intro}

The quest to detect atmospheres on rocky exoplanets requires pushing our observatories to their limits. Even for rocky planets transiting M dwarfs --- the most promising for atmospheric detections --- transmission spectra features have an expected amplitude of $\lesssim$ 20\,ppm. Atmospheric features of such planets are thus perilously close to the pre-launch expected noise floor of JWST instruments ($\sim$~20\,ppm for NIRISS, $\sim$~9\,ppm for NIRCam, and $<$~10\,ppm for NIRSpec; \citealt{Greene2016,Schlawin2021,Rustamkulov2022}). This comparable size of features and noise therefore calls for extra care before claiming an atmospheric detection. In particular, the repeatability of any signal between visits with the same instrument is critical to confirm that observed features are astrophysical in nature.

There has been no definitive or non-disputed detection of an atmosphere on a rocky exoplanet to date. Rocky bodies in the solar system exhibit a ``cosmic shoreline'', which divides bodies with and without atmospheres according to the prevalence of atmospheric escape processes \citep{Zahnle2017}. Such a classification scheme may be a logical starting point for selecting promising rocky exoplanets for atmospheric characterization. However, planets orbiting M dwarfs may exhibit a significantly different cosmic shoreline, or no cosmic shoreline at all, due to the higher stellar activity and extreme-UV (1--912\,\AA) flux levels compared to the Sun, which can strip away planetary atmospheres \citep[e.g.,][]{Owen2012, Becker2020, Dong2020, doAmaral2022}. In fact, recent secondary eclipse observations with the Mid-Infrared Instrument (MIRI) on \jwst of TRAPPIST-1~b and TRAPPIST-1~c, two rocky M-dwarf planets, are consistent with no (or minimal) atmosphere \citep{Greene2023, Zieba2023}.

Several \jwst Cycle 1 programs are surveying rocky planets orbiting M dwarfs, aiming to assess the survivability of their atmospheres (e.g., \jwst\ GO \#1981, 2512, 2589). Through \jwst GO \#1981 (PIs: Stevenson \& Lustig-Yaeger), we are observing five rocky planets around M dwarfs with orbital and planetary properties close to the Solar System's cosmic shoreline. Previous results from this program include two transits of LHS~475~b, which are consistent with no atmosphere or a high-altitude cloud deck (\citetalias{LustigYaeger2023} \citeyear{LustigYaeger2023}), and two transits of GJ~486~b, which are consistent with a H$_2$O-dominated atmosphere or a spectrum contaminated by stellar activity (\citetalias{Moran2023} \citeyear{Moran2023}). Here we present transmission spectra observations for the third planet in the program, \planetname.

\planetname \citep{BertaThompson2015} is a rocky super-Earth (\planetrad, \planetmass, $T_{\rm{eq}}$ = \planettemp; \citealt{Bonfils2018}) orbiting an M dwarf (\starrad, \startemp; \citealt{Bonfils2018}). With a favorable planet-star radius ratio, corresponding to a transit depth of $\sim$0.3\%, \planetname has previously been suggested as a good target for atmospheric characterization \citep[e.g.][]{Schaefer2016}. Assuming a representative atmospheric mean molecular mass of $\mu$ = 10\,AMU, \planetname should have transmission spectra features of $\sim$ 20\,ppm. 

Previous transmission spectra observations of \planetname have a colorful history of claimed atmospheric detections. \cite{Southworth2017} observed nine transits with the MPG 2.2\,m telescope and suggested that deeper transit depths in the $z$ and K bands were caused by a H$_2$-dominated atmosphere with H$_2$O and/or CH$_4$. \cite{DiamondLowe2018} revisited \planetname with five optical transits (0.64--1.04\,$\micron$) with the  Magellan Clay telescope LDSS-3C instrument, finding a featureless transmission spectrum ruling out the previously-claimed atmosphere. \cite{Swain2021} analyzed five near-infrared (1.1--1.7\,$\micron$) Hubble Space Telescope Wide Field Camera 3 (WFC3) transits (\hst GO \#14758, PI: Berta-Thompson), finding evidence for a spectral slope and a feature near 1.53\,$\micron$ suggestive of a H$_2$-dominated atmosphere with aerosols, HCN, and CH$_4$. However, using the same \hst data, \cite{Mugani2021} and \cite{LibbyRoberts2022} do not find these features and instead prefer featureless near-infrared spectra.

In this Letter, we present \jwst transmission spectra observations of \planetname. Our analysis provides a cautionary tale for the challenge of confirming astrophysical signals when searching for rocky exoplanet atmospheres. In Section~\ref{sec:obs}, we describe the observations. In Section~\ref{sec:reduction}, we overview the three independent analysis pipelines used to extract the transmission spectrum. Section~\ref{sec:interpretation} describes our interpretation. Finally, we discuss the implications of these results in Section~\ref{sec:conclusions}.

\section{JWST Observations of \planetname} \label{sec:obs}

We observed two transits of \planetname with the \jwst Near Infrared Spectrograph \citep[NIRSpec,][]{Jakobsen2022,Birkmann2022} G395H instrument on 2023 February 25 and 2023 March 5 as a part of GO \#1981 (PIs: Stevenson \& Lustig-Yaeger). Our data have a spectral resolving power of $\lambda/\Delta\lambda\sim~$2,700 from 2.8--5.2\,$\micron$. Each observation lasted \obstime\,hrs, resulting in \nints integrations each with \ngroups groups up the ramp. The observations were designed to maximize observing efficiency while remaining below the $\sim$80\% full well threshold to avoid the worst impacts of detector non-linearity.

\section{Data Reduction} \label{sec:reduction}

\begin{figure*}[!htb]
    \centering
    \includegraphics[width = 0.95\textwidth]{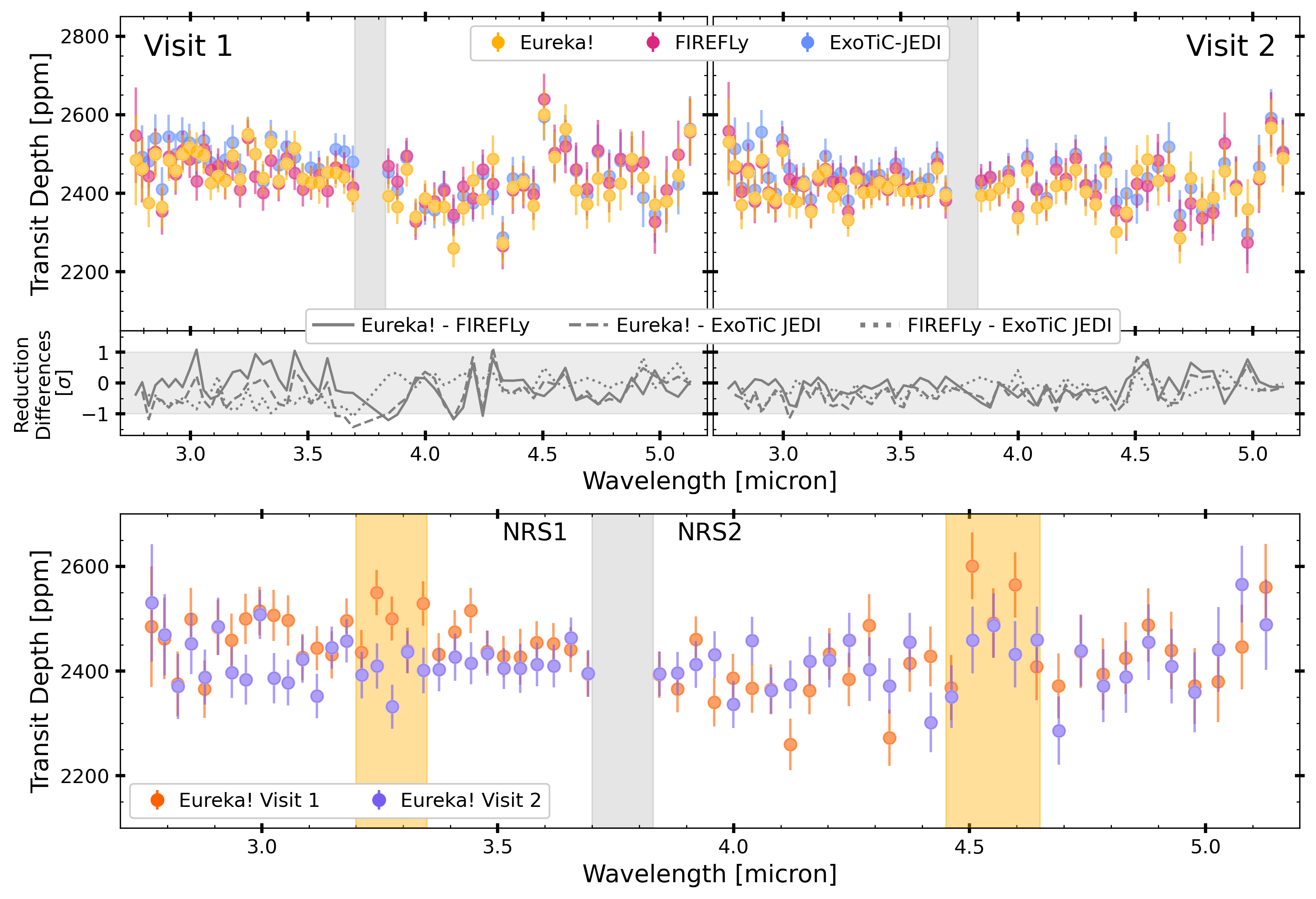}
    \caption{\textbf{Transmission spectra of \planetname from all three data reduction pipelines and both visits.} Upper panel: the \eureka (yellow), \firefly (pink), and \exotic (blue) reductions are shown for both visits at R$\sim$100. Also shown are the differences between reductions (solid, dashed, and dotted gray lines) in units of $\sigma$, demonstrating the agreement between pipelines within a single visit. A $\pm$1$\sigma$ shaded region is overlaid. Bottom panel: the \eureka reduction for both visits, with the NIRSpec detector gap denoted by the shaded gray region. The shaded yellow regions show two important wavelength ranges that differ the most between the visits (see Sections~\ref{ssec:datadiffs} and \ref{ssec:retrievals_atmo} for further discussion on these differences).}
    \label{fig:tspec}
\end{figure*}

To ensure reproducibility of our results, we perform three independent data analyses with the \eureka \citep{Bell2022}, \firefly \citep{Rustamkulov2022, Rustamkulov2023}, and \exotic  \citep{Alderson2022, Alderson2023} pipelines. Previous analyses within \jwst GO \#1981 have shown \eureka and \firefly to agree well for small planetary signals (\citetalias{LustigYaeger2023} \citeyear{LustigYaeger2023}; \citetalias{Moran2023} \citeyear{Moran2023}). While \eureka and \exotic have been shown to agree well for larger planetary signals \citep[e.g.][]{Alderson2023}, here we compare them for signals pushing detection limits. Below, we provide a high-level overview of the data reduction steps from each pipeline. Figure~\ref{fig:tspec} shows the final R$\sim$100 spectra from all three pipelines for both visits. 

\subsection{\texttt{Eureka!}} \label{ssec:eureka}

The \eureka package reduces \jwst time-series data starting from \texttt{uncal} \jwst data through light curve fitting. Stages 1 and 2 of the \eureka pipeline are primarily a wrapper for Stages 1 and 2 of the \texttt{jwst} pipeline \citep{JWST_pipeline}, while also implementing several custom steps, most importantly custom group-level background subtraction. This removes the striping due to 1/f noise at the group level to significantly improve precision (e.g., \citealt{Rustamkulov2022}; \citetalias{LustigYaeger2023} \citeyear{LustigYaeger2023}). This step first identifies the center of light in each column and masks values within 8 pixels on either side. We estimate the background as the mean of all remaining pixels in a column with a 3$\sigma$ outlier threshold. We skip the jump step detection in Stage 1, as we find it only adds noise to the extracted light curves. In Stage 2, we skip the flat field and photom steps, as they result in a conversion to physical flux units that is unnecessary for the relative flux measurements we require. These steps also add noise to the resulting light curves due to the current level of accuracy provided by the available detector flat fields. Additionally, the limited region of the detector that is converted in these steps also worsens the precision on background removal.

Stage 3 of the \eureka pipeline performs spectral extraction after a second round of background subtraction. We use optimal spectral extraction \citep{Horne1986} after correcting the curvature of the trace by measuring the center of light in each column and rolling each column by an integer pixel value to align the trace along the same pixel. The second round of background subtraction considers only the region more than 10 pixels from the trace, with an outlier threshold of 3$\sigma$ in the spatial direction and 10$\sigma$ in the temporal direction. When constructing the median frame for optimal spectral extraction, we employ an outlier rejection threshold of 10$\sigma$ for NRS1 and 15$\sigma$ for NRS2.  For spectral extraction, we use an aperture half-width size of 2 pixels on either side of the center pixel (for a total of 5 pixels), which is selected to achieve the best precision possible by minimizing background noise while maximizing the stellar flux extracted. During spectral extraction, we use an outlier rejection threshold of 7$\sigma$ for NRS1 and 19$\sigma$ for NRS2.  In Stage 4, we generate light curves at the native pixel resolution and at a lower resolution of R$\sim$100.

To determine the best orbital parameters in Stage 5 of \eureka, we first perform joint white light curve fitting across both visits but independently for the two detectors (i.e., NRS1 Visits 1 and 2 are fit jointly). We fit for a linear trend in time combined with a transit function \citep[\texttt{batman};][]{Kreidberg2015}. Limb darkening is fixed to quadratic values obtained with the \texttt{ExoTiC-LD} package \citep{Grant2022} using 3D stellar models \citep{Stagger2012} and assuming stellar values of T$_{\mathrm{eff}}$ = \startemp , $\log{g}$ = 5.02 \citep[both from][]{Stassun2019}, and a metallicity [Fe/H] = -0.12 \citep{BertaThompson2015}. We fit for $R_p/R_\star$, the center of transit, $a/R_\star$, orbital inclination, and the linear term of the temporal ramp. The planet orbital period is fixed to 1.628931 days \citep{Bonfils2018}. We then adopt the weighted mean of the fitted orbital parameters from the NRS1 and NRS2 joint visit fits as our orbital parameter solution. The resulting best fit values are given in Appendix \ref{app:sec:bestfits}, Table \ref{tab:orbital_parameters}, and are held constant in all spectroscopic light curve fits. The only free parameters are then $R_p/R_\star$ and the linear temporal ramp terms for both the native pixel and R$\sim$100 resolutions. All fits use \texttt{emcee} \citep{emcee2013} and are run sufficiently long to ensure chain convergence.\footnote{\eureka control files (\texttt{ecf}) to reproduce these results are available on Zenodo: \href{doi.org/10.5281/zenodo.10002089}{doi.org/10.5281/zenodo.10002089}.}

\subsection{FIREFLy} \label{ssec:firefly}

The \firefly package undertakes the complete process of \jwst time-series data analysis from \texttt{uncal} data files through spectroscopic light curve fitting. We use Stages 1 and 2 of the \texttt{jwst} reduction pipeline for group-level and integration-level detector and instrument corrections, respectively. In Stage 1, the data quality initialization and saturation steps are first applied to the \texttt{uncal} fits files. Unlike the reduction of GJ~486~b (\citetalias{Moran2023} \citeyear{Moran2023}), we do not apply a custom superbias scaling step. In this previous study, we investigated how the bias levels change over the course of the observation. We found that subtracting a scaled superbias file at the group level improved the agreement between the white light curve depths between the two detectors (i.e., helped account for a possible offset between detectors). The scaling factor was calculated by finding the median background per column in each group, dividing this by the median per column superbias value, and averaging all columns to get a single scaling factor per group. For GJ1132b, we instead elect to use the single default \texttt{jwst} superbias file, as this results in the most consistent white light curve transit depths between the four datasets (NRS1 and NRS2 for Visits 1 and 2). We do, however, implement a custom background subtraction to reduce 1/f noise at the group-level in Stage 1. We then apply the reference pixel correction and linearity step while skipping the dark current step. We also skip the jump step, which is only applied in the \firefly pipeline if the number of groups per integration is larger than 25, as fewer groups per integration (here, 14) lowers the risk of cosmic ray hits. After ramp fitting and the gain-scale step, in Stage 2 we use only the assign WCS step (skipping the flat field step) before proceeding to \firefly's stellar extraction.

For stellar extraction, we first clean bad pixels using \texttt{lacosmic} \citep{lacosmic}. We determined the bad pixel map by flagging pixels with sharp variance spikes and manually checking known bad pixels in NIRSpec G395H. We apply another background subtraction at the integration level, measure the x- and y-shifts, and finally extract the 1D stellar spectrum. We use a pre-calculated trace and aperture full-width of 5.93 pixels (optimized from previous NIRSpec G395H observations; e.g., \citetalias{LustigYaeger2023} \citeyear{LustigYaeger2023}, \citetalias{Moran2023} \citeyear{Moran2023}) and compute a box extraction.

We fit \planetname's light curves with \texttt{batman} \citep{Kreidberg2015}, both at the native pixel resolution and at R$\sim$100. For the R$\sim$100 case, we trim the first 150 columns of NRS1. To fit the white light curve, we fix the orbital period to 1.629 days \citep{Bonfils2018} and fit for $R_p/R_{\star}$, the mid-transit time $T_0$, $a/R_{\star}$, the impact parameter $b$, the quadratic limb darkening coefficients, and several systematics parameters. We fit each of the four datasets individually, using the Bayesian Information Criterion (BIC) to determine the best-fit systematics model in each case. Therefore, different models can be used for different datasets. Specifically, we use a linear ramp in time for NRS1 (both visits) and NRS2 Visit 1. For NRS2 Visit 2, we use only the y-shift. After light curve  fitting using these optimized systematics, we refit all four datasets using a weighted average of $a/R_{\star}$, $b$, and the limb darkening coefficients. 

To extract the transmission spectrum, we fit the spectroscopic light curves (holding the orbital parameters and limb darkening coefficients fixed to their white light curve values) with only $R_P/R_{\star}$ and the corresponding systematics as free parameters. While \firefly utilizes \texttt{emcee} \citep{emcee2013} when fitting the white light curve, the spectroscopic fits use least-squares fitting with \texttt{lmfit} \citep{Newville2014}. This method is much faster and does not impact the resulting spectrum. We extensively compared both techniques at the native pixel level and at R$\sim$100 and found no meaningful changes when using least-squares fitting. 

\subsection{ExoTiC-JEDI} \label{ssec:exotic}

The Exoplanet Timeseries Characterisation - JWST Extraction and Diagnostic Investigator (\exotic) package\footnote{https://github.com/Exo-TiC/ExoTiC-JEDI} performs a full extraction, reduction, and analysis of \jwst time-series data from \jwst \texttt{uncal} files to light curve fitting. For NIRSpec G395H observations, we treat the NRS1 and NRS2 datasets independently. We use the \texttt{jwst} pipeline tools to perform linearity, dark current, and saturation corrections, with the jump detection threshold set to 15, and a custom destriping routine to remove 1/f noise at the group level using a second-order polynomial and a threshold of 15$\sigma$ fit to the background region. These steps are followed by a standard ramp fitting routine. \exotic is also able to perform custom bias subtraction, but we find that it does not improve the precision of the data in this case. We extract our Stage 2 products, the 2D wavelength array, and exposure times using the standard \texttt{jwst} pipeline. 

Stage 3 of the \exotic package performs pixel corrections, additional background and 1/f noise removal, and spectral extraction. Using the data quality flags provided from the \texttt{jwst} pipeline, we replace pixels identified as bad, saturated, low quantum yield, hot, dead, or no gain with the median of the surrounding pixels. To remove additional bad pixels due to cosmic rays or other phenomena, we identify both spatial and time-series outliers in the data cube with a 20$\sigma$ threshold in time and 6$\sigma$ threshold spatially, replacing any identified pixels with the median of that pixel in the surrounding 10 integrations or 20 pixels in that row. Any remaining 1/f noise is removed by masking the illuminated region of the detector to calculate the median illuminated pixel value in each column. To extract the 1D stellar spectrum, we fit a Gaussian to each column of the data followed by a fourth-order polynomial to the trace center and widths ($\sim$ 0.7 pixels wide). The trace centers and widths are then smoothed with a median filter and used to determine a simple aperture region 5$\times$ the trace FWHM ($\approx$ 7 pixels). We use intrapixel extraction to obtain our 1D stellar spectrum. At Stage 3, we also measure the trace position movement on the detector in the x- and y-position for detrending at later stages.  

We perform light curve fitting on broadband NRS1 and NRS2 spectra, as well as spectroscopically across the full wavelength range. Using the broadband spectra for each detector and visit, we fit for the planetary system inclination and $a/R_\star$ while fixing the period (1.628931\,days) and eccentricity (0.0) to literature values presented by \citet{Bonfils2018}; as the eccentricity value presented in \citep{Bonfils2018} is an upper limit we test the fit fixed at both e=0.0 and e=0.22 and find no impact on the resultant transmission spectrum. These parameters, along with the center of transit time, are held constant in the spectroscopic light curve analysis. Stellar limb-darkening coefficients are calculated using the \texttt{ExoTiC-LD} package with a custom model input using a Phoenix stellar model \citep[][T$\mathrm{eff}$=3300\,K, logg=5.0, $\mathrm{[Fe/H]}$=0.0]{Husser2013} and the nonlinear limb-darkening law. Limb darkening values are then fixed in our light curve analysis. We use a least-squares optimizer with a \texttt{batman} \citep{Kreidberg2015} transit model to fit for the transit depth in each bin. We simultaneously fit a series of systematic models to the data and determine the optimal model based on the negative log-likelihood, which incorporates a penalization in complexity based on the AIC (Akaike Information Criterion). We find that the best systematic model, $S(\lambda)$, corrects for a linear trend in time, $t$, plus the change in x-position, $x_{s}$, multiplied by the absolute magnitude of the y-positional change, $|y_{s}|$, such that $S(\lambda) = s0 + (s1 \times x_{s}|y_{s}|) + (s2 \times t) \mathrm{,}$ (where $s0, s1, s2$ are coefficient terms). 

\subsection{Agreement between Pipelines and Resolutions}

Figure~\ref{fig:tspec} shows the R$\sim$100 transmission spectra for all three reductions and both visits, demonstrating excellent agreement between the pipelines. We find that our three R$\sim$100 reductions agree better than the native resolution light curve fits, particularly at the red end of the spectrum where the signal-to-noise (SNR) decreases due to a combination of decreased throughput and decreased stellar signal (not shown here). While previous work has found that noise is reduced by performing light curve fits at native pixel resolution prior to spectrally binning the data  \citep{Espinoza2023}, our results suggest that low-SNR transit signals can impart biases that can change the shape of the spectrum when fit at native pixel resolution. \citetalias{LustigYaeger2023} (\citeyear{LustigYaeger2023}) showed that the precision improvement from fitting at the native pixel resolution can be negated by sufficiently removing 1/f noise at the group level, suggesting that binning light curves prior to fitting is acceptable. Our results here show that binning light curves prior to fitting may be preferable for low-SNR targets (specifically when the spectrophotometric scatter approaches the transit depth). We further discuss the impact of native resolution light curve fitting in Section~\ref{sec:binning}.

\begin{deluxetable*}{l||l|r|c||l|r|c||l|r|c}
\tablewidth{0.98\textwidth}
\tabletypesize{\footnotesize}
\tablecaption{Is it Flat? \label{tab:flat}}
\tablehead{\colhead{} & \multicolumn{3}{c}{\bf Visit 1} & \multicolumn{3}{c}{\bf Visit 2} & \multicolumn{3}{c}{\bf Combined}  \\ 
\colhead{Reduction} & \colhead{$\chi^{2}_{\nu}$} & \colhead{p-value ($\sigma$)} & \colhead{Feature?} & \colhead{$\chi^{2}_{\nu}$} & \colhead{p-value ($\sigma$)} & \colhead{Feature?} & \colhead{$\chi^{2}_{\nu}$} & \colhead{p-value ($\sigma$)} & \colhead{Feature?}}
\startdata
\eureka  & 1.57 & 0.32\% (2.9$\sigma$) & 3.4$\sigma$  & 0.78 & 89\% (0.13$\sigma$) & -2.5$\sigma$ & 1.48 & 0.92\% (2.6$\sigma$) & 1.0$\sigma$  \\
\firefly & 1.08 & 32\% (1.0$\sigma$)   & -2.4$\sigma$ & 0.76 & 91\% (0.11$\sigma$) & -2.7$\sigma$ & 1.14 & 21\% (1.3$\sigma$)   & -2.4$\sigma$ \\ 
\exotic  & 1.54 & 0.47\% (2.8$\sigma$) & 4.3$\sigma$  & 0.89 & 72\% (0.36$\sigma$) & -2.6$\sigma$ & 1.50 & 0.84\% (2.6$\sigma$) & 3.8$\sigma$  \\  
\hline
\eureka\tablenotemark{a}  &  1.23 & 10.5\% (1.6$\sigma$) & 1.4$\sigma$  & 0.78 & 90\% (0.13$\sigma$) & -2.3$\sigma$ & 1.27 & 7.7\% (1.8$\sigma$) & -1.0$\sigma$ \\
\firefly\tablenotemark{a} &  1.02 & 44\% (0.77$\sigma$)  & -2.0$\sigma$ & 0.76 & 92\% (0.10$\sigma$) & -2.6$\sigma$ & 1.11 & 26\% (1.1$\sigma$)  & -2.4$\sigma$ \\ 
\exotic\tablenotemark{a}  &  1.03 & 42\% (0.8$\sigma$)   & -1.8$\sigma$ & 0.86 & 76\% (0.30$\sigma$) & -2.5$\sigma$ & 1.12 & 25\% (1.1$\sigma$)  & -2.3$\sigma$ \\ 
\enddata
\tablecomments{$\chi^{2}_{\nu}$ is the reduced chi-squared resulting from the best fitting featureless fit (null hypothesis) to the observed spectrum, ``p-value'' refers to the probability that $\chi^{2}_{\nu}$ would be at least as extreme as the observed value under the assumption that the null hypothesis is correct and is displayed along with the corresponding ``sigma'' value, and the ``Feature?'' column shows the level of confidence in the detection of an agnostic Gaussian absorption feature in the spectrum over the null hypothesis (negative values denote preference for the featureless model).} 
\tablenotetext{a}{Allows NRS2 data to shift relative to NRS1.}
\end{deluxetable*}

\section{Interpretation} \label{sec:interpretation}

To interpret our \planetname transmission spectra, we first explore the consistency between visits then conduct stellar and planetary atmosphere forward modeling and retrievals on both visits independently. 

\subsection{Differences Between Visits} \label{ssec:datadiffs}

While the data reduction pipelines show good agreement, the transmission spectra show notable differences between Visit 1 and Visit 2 (see Figure \ref{fig:tspec}). We begin our analysis by investigating the statistical significance of these differences against the null hypothesis of a flat, featureless transmission spectrum. We approach this in two ways, as described below.

First, we assess how expected or unexpected our measurements would be under the assumption that the spectrum is featureless. We report in Table~\ref{tab:flat} the reduced chi-squared, $\chi^{2}_{\nu}$, for the featureless spectrum (flat line model) that best fits each dataset. We then calculate the distribution of expected $\chi^{2}_{\nu}$, under the assumption that the null hypothesis is true, using 100,000 randomly generated synthetic featureless spectra with the same uncertainties as the observed data. We calculate the probability that $\chi^{2}_{\nu}$ would be at least as extreme as the observed value under the assumption that the null hypothesis is correct (i.e., the ``p-value''). Table~\ref{tab:flat} reports the p-values and corresponding ``sigma'' significance with which the null hypothesis is disfavored by the test. 

Second, following \citetalias{Moran2023} (\citeyear{Moran2023}), we fit each spectrum with a Gaussian model (representing an agnostic spectral feature) and compare it to the featureless model. Both the Gaussian and featureless models are fit using the \texttt{Dynesty} nested sampling code \citep{Speagle2020}, which returns the Bayesian evidence for each fit. From the evidence, we calculate a Bayes factor and convert it into a ``sigma'' value \citep{Trotta2008} representing the significance of the Gaussian feature model over the featureless model. These results are reported in Table~\ref{tab:flat}, where positive values denote evidence favoring the Gaussian model and negative values favor the featureless model. We note that these results can differ from the first statistical test, depending on how well a single Gaussian with varying wavelength center, width, and amplitude can actually fit the spectrum.  

The conclusions from our two statistical tests are consistent, despite having slightly different numerical results. In general, Visit 1 contains marginal evidence to reject the null hypothesis and favor a non-flat spectrum, while Visit 2 is statistically consistent with a flat line. For Visit 1, the \firefly reduction has a $\chi^{2}_{\nu}$ near unity (indicating it favors the null hypothesis), which disagrees with that of \eureka and \exotic --- likely owing to the slightly larger uncertainties in the \firefly reduction. However, all three reductions agree well for Visit 2 and favor a featureless spectrum.

We also investigated the sensitivity of our results to a potential transit depth offset between the NRS1 and NRS2 detectors for NIRSpec G395H. The second set of rows in Table~\ref{tab:flat} show the results for the same tests as the top three rows, but now allowing for NRS2 to shift vertically relative to NRS1 to account for a potential systematic offset between the two detectors. These results show that including an offset for NRS2 erodes the statistical significance with which to reject the null hypothesis for Visit 1, while leaving the results for Visit 2 largely unchanged. In Appendix~\ref{app:sec:gaussians}, Figure~\ref{app:fig:flat}, we provide a visual example of these null hypothesis tests. These results demonstrate that the two visits are inconsistent unless we allow for a transit depth offset between the two detectors. There is, however, no strong evidence for the necessity of a significant detector offset, since there was no need for a superbias correction step (unlike for GJ~486~b, where this was the primary driver for a detector offset, see \citetalias{Moran2023} \citeyear{Moran2023}). Even when including an offset, Visit 1 is still more consistent with spectral features than Visit 2.

\subsection{Evidence for Variable Star Heterogeneity?} \label{ssec:stellar}

A natural physical explanation for different transmission spectra for a planet orbiting an M dwarf is stellar variability \citep[e.g.,][]{Rackham2018}. To investigate the possibility of changing starspot coverage, we first perform forward modeling of the out-of-transit extracted, flux-calibrated stellar spectra for both visits.

We compute multi-component stellar forward models using the \cite{Allard2012} \texttt{PHOENIX} models. We seek plausible evidence of evolution or rotation of features onto or off of the observed disk by quantifying the spot and faculae covering fractions of the stellar surface for Visits 1 and 2. We followed a similar procedure to  \citetalias{Moran2023} (\citeyear{Moran2023}), employing a weighted linear combination of three \texttt{PHOENIX} models to represent the background photosphere, spots ($T_{\rm eff}$ $\leq$ $T_{\rm eff, \, photosphere}$ - 100\,K), and faculae ($T_{\rm eff}$ $\geq$ $T_{\rm eff, \, photosphere}$ + 100\,K). We assume that all spots have a common $T_{\rm eff}$, log($g$), and metallicity (as do the faculae), and constrain each feature to not exceed 45\% of the stellar surface. The grid of \texttt{PHOENIX} models used for our analysis covers $T_{\rm eff}$ = 2500--4500\,K, log($g$) = 4--5.5\,cm\,s$^{-2}$, and [Fe/H] = -0.5--0, which provides extensive coverage of possible spot and faculae temperatures for GJ~1132. Finally, we assume photospheric values near the literature quoted $T_{\rm eff}$ = 3270 $\pm$ 140\,K; \citep{Bonfils2018}, log($g$) = 4.88 $\pm$ 0.07\,cm\,s$^{-2}$ \citep{Southworth2017}, and [Fe/H] = -0.12 $\pm$ 0.15 \citep{BertaThompson2015}.

To compare with the observed baseline spectra, we first convert the native model wavelengths (\AA{}) and flux densities (erg\,s$^{-1}$\,cm$^{-2}$\,cm$^{-1}$) to $\micron$ and mJy. We also scaled the models by R$_*^2$/$d^2$ using literature values for GJ~1132: R$_\star$=0.21\,R$_\odot$ \citep{Bonfils2018} and $d$ = 12.61\,pc \citep{Gaia2021}. We smoothed and interpolated the models the same resolution as the observations before calculating a $\chi_\nu^2$. In our $\chi_\nu^2$ calculations, we considered 3206 wavelength points for each visit and eight fitted parameters (the $T_{\rm eff}$, log($g$), and [Fe/H] of the photosphere, the $T_{\rm eff}$ and coverage fraction of both spots and faculae, and a scaling factor). The scaling factor was multiplied by the R$_\star^2$/$d^2$ term to account for uncertainty in either measured quantity and varied from 0.9 to 1.1.

\begin{figure*}[!htb]
    \centering
    \includegraphics[trim = 0.0cm 0.0cm 0.0cm 0.0cm, width = 1.0\textwidth]{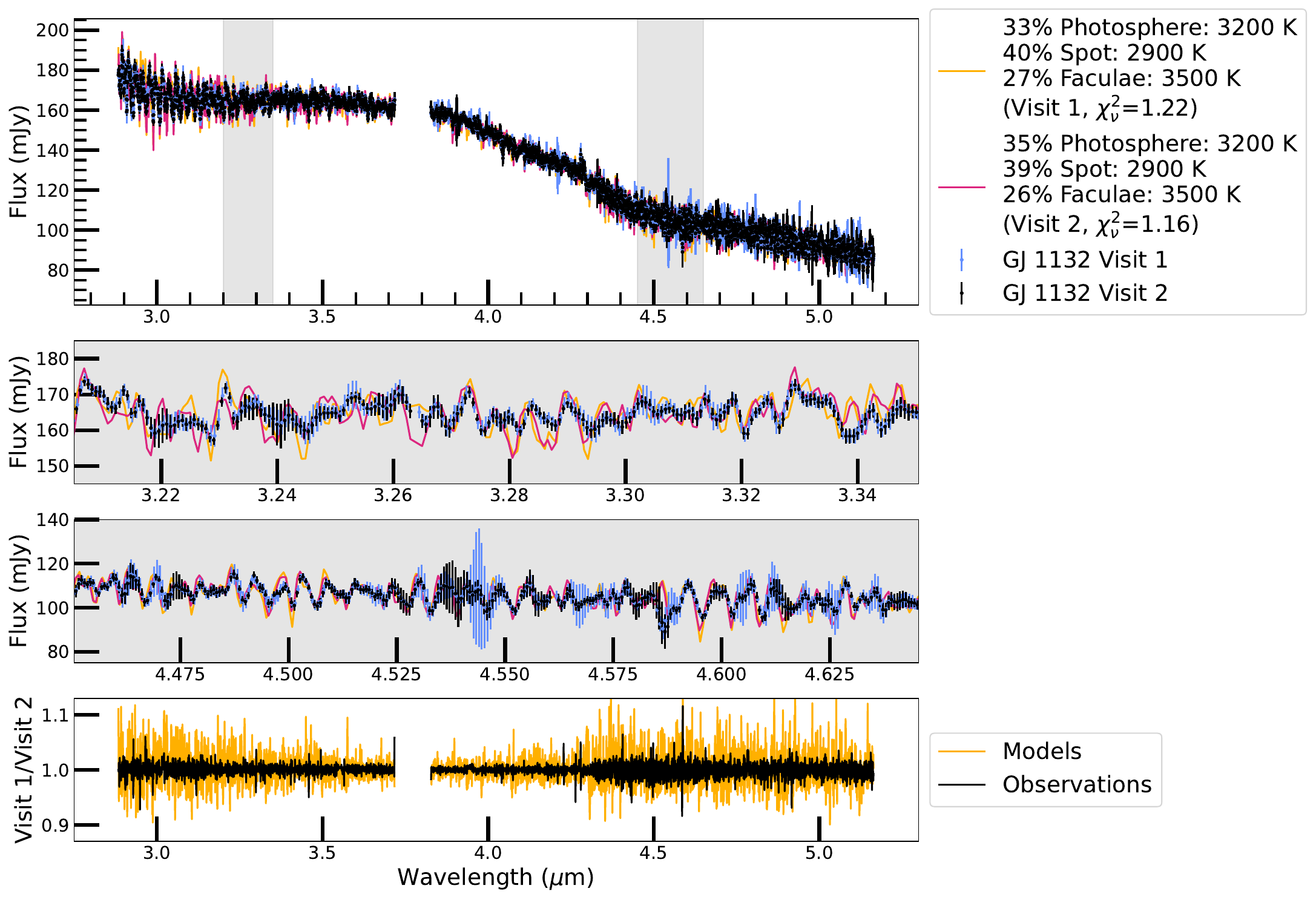}
    \caption{\textbf{Out-of-transit stellar spectra of GJ~1132 compared to heterogeneous stellar models}. Top panel: the extracted GJ~1132 spectra from Visit 1 (blue errors) and Visit 2 (black errors) are compared to best-fitting multi-component \texttt{PHOENIX} models (yellow for Visit 1; pink for Visit 2). The best-fitting Visit 1 model has 40\% spot coverage and 27\% faculae coverage, compared to 39\% spot coverage and 26\% faculae coverage for Visit 2. Middle panels: zoom-ins on the gray highlighted regions in the top panel. Bottom panel: deviations between visits for the models (yellow) and observations (black). The largest deviations for both models and observations occur near 3\,$\micron$ and long-ward of 4.3\,$\micron$.}
    \label{fig:phoenix}
\end{figure*}

Our out-of-transit stellar spectra and best-fitting models for both visits are shown in Figure~\ref{fig:phoenix}. The preferred models for Visit 1 and Visit 2 both have a background photosphere with $T_{\rm eff}$ = 3200\,K, log($g$) = 4.5 cgs, and [Fe/H] = 0, spots with $T_{\rm eff}$ = 2900\,K, and faculae with $T_{\rm eff}$ = 3500\,K (top panels, Figure~\ref{fig:phoenix}). The preferred model for Visit 1 is 33\% photosphere, 40\% spots, and 27\% faculae ($\chi_\nu^2 = 1.22$). The preferred model for Visit 2 is 35\% photosphere, 39\% spots and 26\% faculae (with a $\chi_\nu^2$ of 1.16). The largest spectral deviations between visits occur from 2.75--3.3\,$\micron$ and 4.3--5.3\,$\micron$ for both the observations and the models (bottom panel, Figure~\ref{fig:phoenix}). However, a 1\% difference in spot and faculae coverage is negligible, considering general model uncertainties and error bars on the observations. Therefore, it is unlikely that the differences between visits are caused by the evolution of surface features or rotation onto or off of the visible disk.

We further note that there is no evidence for occulted starspots in either of the visits, which could otherwise explain the different transmission spectra due to the impacts on the morphology of the transit light curves themselves. Figure~\ref{app:fig:2d} in Appendix~\ref{app:sec:2d} shows the \eureka white light curves for both visits, showing the lack of obvious spot occultations in either visit (occulted starspots would result in a brief decrease in transit depth during the transit event). However, because occulted starspots typically have a minimal impact on the visible shape of a light curve at these wavelengths, we also consider how the fitted orbital parameters may change due to such features. \firefly and \exotic fit the white light curves from the two visits independently, and while the independent fit values are not reported here, they are consistent within uncertainties.

\subsection{An Atmosphere Around \planetname?} \label{ssec:atmo}

\begin{figure*}[!htb]
    \centering
    \includegraphics[trim = 0.0cm 0.0cm 0.0cm 0.0cm, width = 1.0\textwidth]{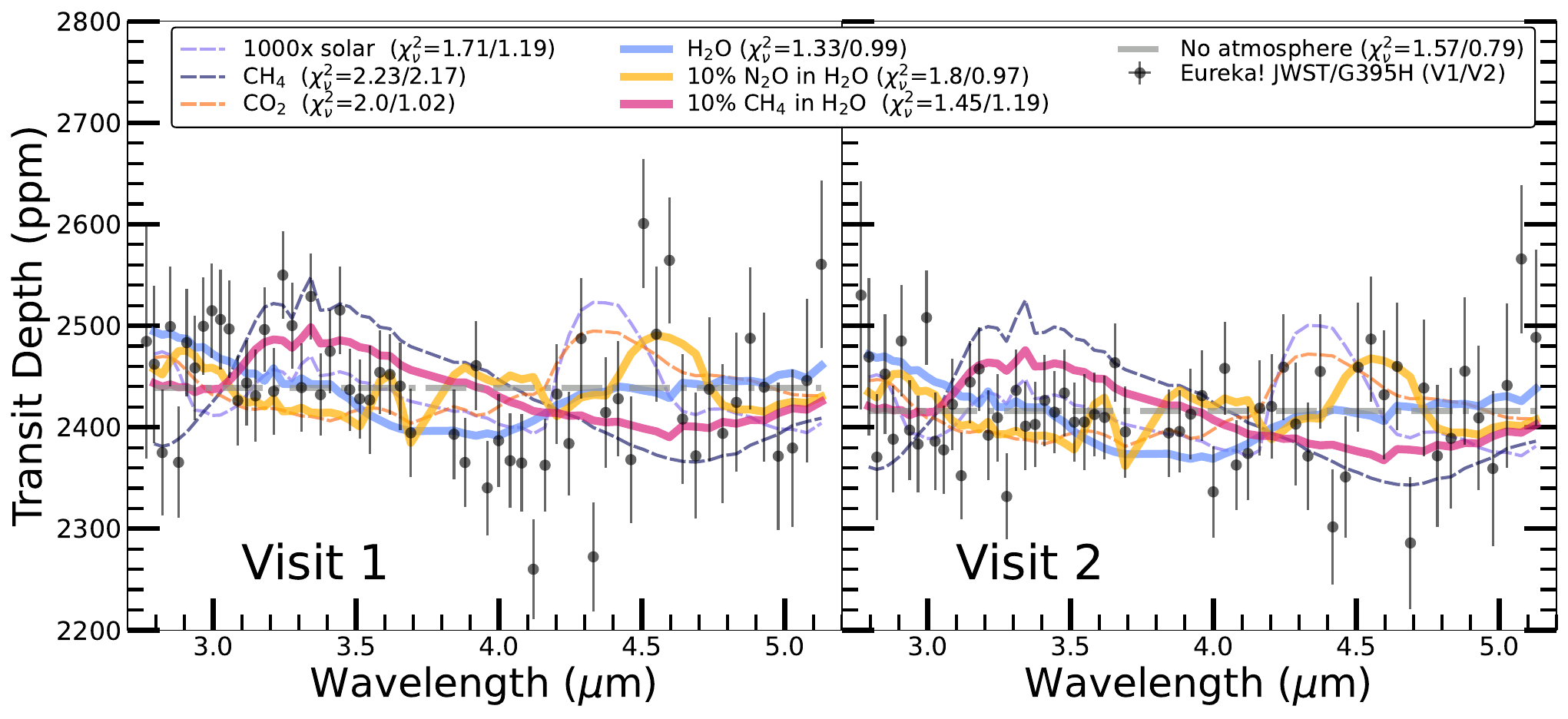}
    \caption{\textbf{Transmission spectra of \planetname compared to atmospheric forward models.} The \eureka reduction (black circles) is shown for both visits at an R$\sim$100, with Visit 1 on the left and Visit 2 on the right. Also shown are a series of end-member atmospheric forward models generated with \texttt{PICASO} to illustrate goodness of fit ($\chi$$^2$$_\nu$) of various scenarios for each visit, shown after each model for Visit 1 and 2, respectively. Dashed lines show poorer overall fits including a 1000$\times$ solar atmosphere (dashed purple), a pure 1 bar CH$_4$ atmosphere (dashed navy), and a pure 1 bar CO$_2$ atmosphere (dashed orange). Better fits include pure H$_2$O atmospheres (blue) or water atmospheres with 10\% N$_2$O (yellow) or 10\% CH$_4$ (pink), or a flat line indicative of no atmosphere or a high altitude opaque aerosol layer (gray dashed). In Visit 1, water-rich atmospheres with other species are preferred, but the features driving this fit disappear in Visit 2 so that an atmosphere-free model is the best fit.}
    \label{fig:picaso}
\end{figure*}

We next assess possible atmospheric explanations for \planetname's transmission spectrum. As in our previous studies (\citetalias{LustigYaeger2023} \citeyear{LustigYaeger2023}; \citetalias{Moran2023} \citeyear{Moran2023}), we compare each reduction to a set of simple forward models to compare possible atmospheric compositions and a no-atmosphere scenario. We first generate forward model atmospheres using either thermochemical equilibrium \texttt{CHIMERA} \citep{line2013,line2014} models or simple one-or-two gas isothermal atmosphere \texttt{PICASO} \citep{Batalha2019} models. We then compute  model transmission spectra using \texttt{PICASO}'s radiative transfer module. We bin the resulting model spectra to the resolution of each reduction and compute a $\chi_\nu^2$ (with 60 degrees of freedom (dof) for the \eureka and \firefly reductions and 59 for the \exotic reduction) to assess goodness of fit. We summarize our results in Figure~\ref{fig:picaso}.

First, we run a set of thermochemical equilibrium models with \texttt{CHIMERA}. We include opacities from H, collision-induced absorption (CIA), H$_2$, He, H$_2$O, CH$_4$, CO, CO$_2$, NH$_3$, N$_2$, HCN, and H$_2$S and use the parameterized temperature-pressure profile of \citet{guillot2010} with an equilibrium temperature of 530\,K. We run these forward models at 100$\times$ to 1000$\times$ solar metallicities, finding that the scale height of the atmosphere is so large in all cases that we obtain poor fits for all reductions in Visit 1 ($\chi_\nu^2$ $\gtrsim$ 1.43). Therefore, we rule out clear hydrogen-dominated atmospheres in thermochemical equilibrium at moderate confidence ($\gtrsim$2.5$\sigma$). For Visit 2, these confidences are decreased for the 1000$\times$ solar metallicity case (down to 1.3$\sigma$ for the {\firefly} reduction), but a hydrogen-dominated atmosphere is never the statistically preferred scenario compared to our forward models. A long-lived hydrogen-dominated atmosphere would be unexpected given the planet's density and radius  \citep{luger2015,rogers2015,Estrela2020,Rogers2021}, so our forward model limits here fit expectations better than some previous results for \planetname \citep{Southworth2017,Swain2021}.

For our simpler, non-self-consistent \texttt{PICASO} models, we examine whether 1\,bar, isothermal atmospheres of pure CH$_4$, pure CO$_2$, or pure H$_2$O agree with the data from each reduction. We find that CH$_4$-dominated atmospheres (dark blue dashed line in Figure~\ref{fig:picaso}) are the most strongly ruled out for both visits, to at least 4.2$\sigma$ across all reductions. Although in Visit 1 there is a rise in transit depth at the strong CH$_4$ absorption feature centered at 3.3\,$\micron$, the lack of strong CH$_4$ absorption at the wavelengths probed by NRS2 result in a poor fit. Similarly, CO$_2$-dominated atmospheres poorly fit Visit 1. However, H$_2$O-rich atmospheres (thick solid lines in Figure~\ref{fig:picaso}) provide better fits to Visit 1 due to the broad H$_2$O absorption slope at the bluest wavelengths probed by NIRSpec G395H. In Visit 1, an uptick in transit depth at $\sim$ 4.5\,$\micron$ is also noticeable. Multiple molecules, including O$_3$, CS$_2$, and N$_2$O, have an absorption band around this wavelength \citep[e.g.,][]{Schwieterman2022}, but of these N$_2$O has the best-matching feature center and width. Therefore, in addition to our pure H$_2$O atmosphere, we also generate atmospheric models with H$_2$O as the background gas and either 10\% N$_2$O or 10\% CH$_4$. These result in visually better fits at the two increases in transit depth at 3.3\,$\micron$ and 4.5\,$\micron$. A flat-line fit is slightly disfavored for Visit 1 (rejected between 1-3$\sigma$) --- consistent with our previous statistical tests in Section~\ref{ssec:datadiffs} --- in favor of H$_2$O-rich atmospheres.

In summary, our forward models prefer a H$_2$O-dominated atmosphere for Visit 1. While adding 10\% CH$_4$ or N$_2$O visually explain the observed features, they do not improve $\chi_\nu^2$ (but we only consider a single mixing ratio for both species). We explore the full range of possible mixing ratios consistent with Visit 1 with atmospheric retrievals in Section~\ref{ssec:retrievals}.

For Visit 2, we find that a flat line --- indicative of either no atmosphere or a high altitude aerosol layer --- produces the lowest $\chi_\nu^2$ (consistent with Section~\ref{ssec:datadiffs}). At \planetname's $\sim$500\,K equilibrium temperature, while condensate clouds are unlikely (given the lack of condensible species) photochemical hazes could form in a variety of atmospheres \citep[e.g.,][]{Horst2018,He2018,Gao2020}. Moreover, we cannot rule out the H$_2$O-dominated atmosphere preferred by Visit 1 from the Visit 2 data, given its low $\chi_\nu^2$ ($<$ 1).

\begin{figure*}[ht!]
    \centering
    \includegraphics[width=\textwidth]{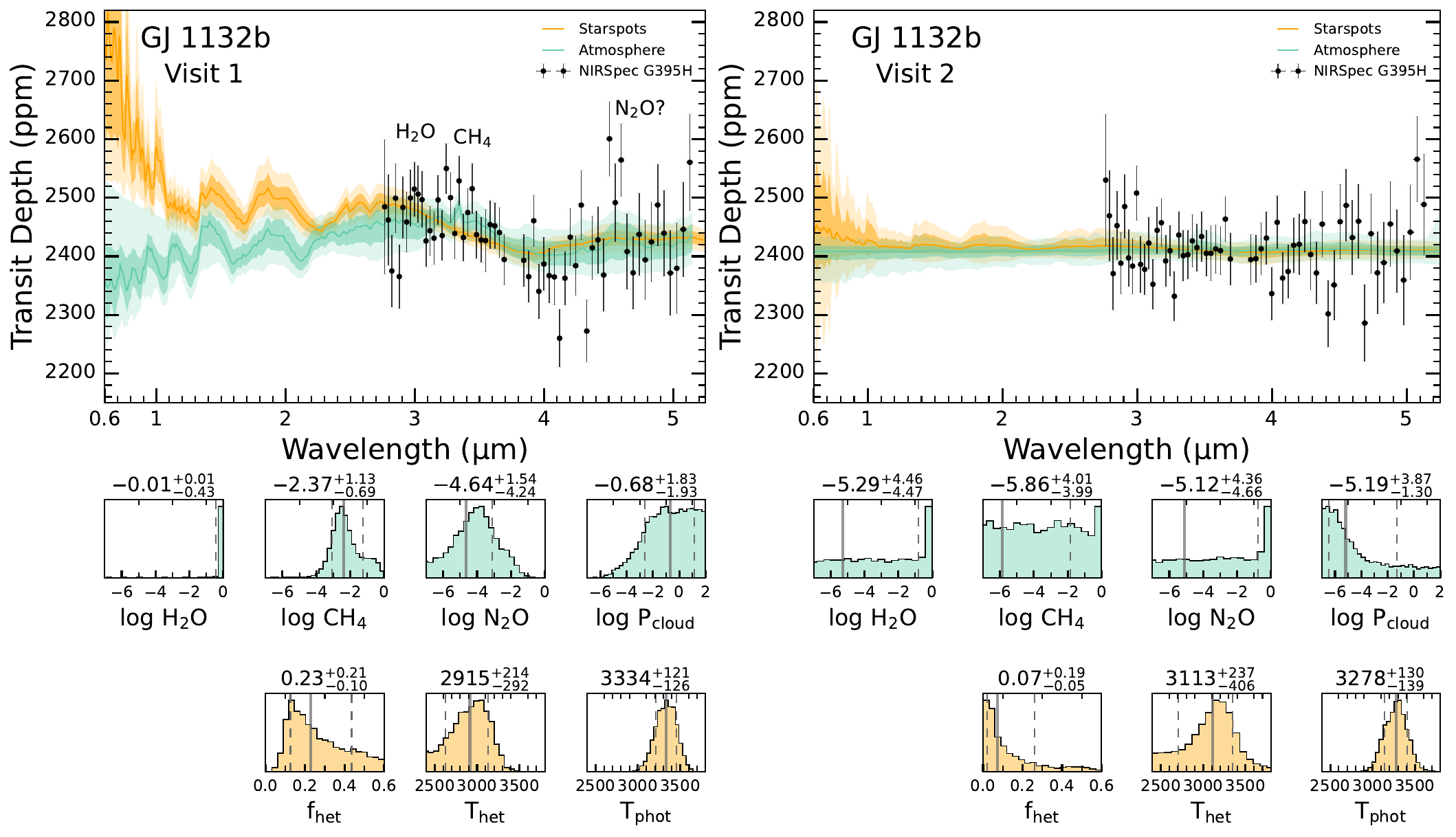}
    \caption{\textbf{Atmospheric and starspot retrieval results for \planetname.} Top panels: comparison between the retrieved transmission spectra for the \eureka reduction of Visit 1 (left) and Visit 2 (right) adopting two distinct retrieval models: (i) a planetary atmosphere with no unocculted starspots (green contours), and (ii) no atmosphere with unocculted starspots (orange contours). The median retrieved spectrum (solid lines) and 1$\sigma$ and 2$\sigma$ model confidence intervals (dark and light contours) for each scenario are overlaid. Labels indicate the locations of H$_2$O, CH$_4$, and N$_2$O absorption bands. Middle panels: posterior histograms for the atmosphere scenario, highlighting the volume mixing ratios of the three molecules tentatively inferred from Visit 1: H$_2$O, CH$_4$, and N$_2$O. Bottom panels: posterior histograms for the unocculted starspot scenario, defined by the spot coverage fraction, spot temperature, and the background stellar photosphere temperature. Visit 1 requires either a water-rich atmosphere with trace CH$_4$ (and possibly N$_2$O) or unocculted starspots, but the scenarios cannot be differentiated without observations shortwards of 3\,$\micron$. Visit 2 is, however, consistent with no atmosphere and no unocculted starspots.}
     \label{fig:retrievals}
\end{figure*}

\subsection{Retrieval Analysis: An Atmosphere or Starspots?} \label{ssec:retrievals}

Our analysis thus far has yielded two key results: (i) GJ~1132's out-of-transit stellar spectrum strongly favors a heterogeneous star with constant spot and faculae properties between the two Visits, and (ii) \planetname's Visit 1 spectrum can be explained by an atmosphere, but Visit 2 is statistically flat. Here we attempt to reconcile these insights via retrieval modeling of \planetname's transmission spectrum considering both atmospheric and unocculted starspot scenarios. Our retrieval results are summarized in Figure~\ref{fig:retrievals}.

\subsubsection{Atmosphere Scenario} \label{ssec:retrievals_atmo}

We first explore the range of atmospheres consistent with \planetname's transmission spectrum via separate retrievals of our two visits using the open source \POSEIDON code \citep{MacDonald2017,MacDonald2023}. We considered 11 potential gases to span a wide parameter space of plausible atmospheric compositions: N$_2$, H$_2$, H$_2$O, CO$_2$, CH$_4$, N$_2$O, NO$_2$, HCN, NH$_3$, SO$_2$, and PH$_3$. The opacities used for the retrieval forward model are described in \citet{MacDonald2022}. The mixing ratios of these gases can range from $10^{-12}$ to 1, using centered log-ratio priors as in \citetalias{LustigYaeger2023} (\citeyear{LustigYaeger2023}) and \citetalias{Moran2023} (\citeyear{Moran2023}). The other free parameters are (priors in brackets) the atmospheric temperature ($\mathcal{U}$(400\,K, 900\,K)), the atmosphere radius at the 10\,bar reference pressure ($\mathcal{U}$(0.85\,$R_{\rm{p, \, obs}}$, 1.15\,$R_{\rm{p, \, obs}}$)), the haze power-law exponent ($\mathcal{U}$(-20, 2)) and log-Rayleigh enhancement factor ($\mathcal{U}$(-4, 8)) --- defined as in \citet{MacDonald2017} --- and the log-pressure of an opaque cloud/surface ($\mathcal{U}$(-7, 2), in bar). We calculate transmission spectra via opacity sampling at a resolving power of $R =$ 20,000 from 0.6--5.2\,$\micron$, before convolving the model with the instrument point spread function and binning to the resolution of the observations. We sample this 15-parameter space using the \texttt{PyMultiNest} \citep{Feroz2009,Buchner2014} package with 2,000 live points.

We perform retrievals on each \planetname visit separately to consider how their different morphology (discussed in Section \ref{ssec:datadiffs}) affects atmospheric inferences. Our retrievals here focus on the \eureka reduction since we found consistent retrieval results with \exotic and \firefly. We do not consider a free detector offset between the NRS1 and NRS2 spectra in these retrievals, given the lack of evidence for a superbias correction during the data reduction.

Our Visit 1 retrieval (left panels of Figure \ref{fig:retrievals}) favors a H$_2$O-dominated atmosphere (Bayes factor = 6.5 / 2.5$\sigma$) with trace amounts of CH$_4$ (Bayes factor = 27 / 3.1$\sigma$). Figure~\ref{fig:retrievals} demonstrates that the evidence for H$_2$O is driven by the slope seen in the NRS1 data, while a feature near 3.3\,$\micron$ is attributed to CH$_4$. A weak feature near 4.5\,$\micron$ is best fit by N$_2$O absorption, but our Bayesian model comparison indicates insufficient evidence for N$_2$O (Bayes factor = 1.2 / 1.4$\sigma$). The volume mixing ratio posterior distributions in Figure~\ref{fig:retrievals} show a H$_2$O abundance consistent with 100\% ($\log X_{\mathrm{H_2 O}} = -0.01^{+0.01}_{-0.43}$), a CH$_4$ abundance of $\sim$ 1\% ($\log X_{\mathrm{CH_4}} = -2.37^{+1.13}_{-0.69}$), and an N$_2$O abundance of $\sim$ 100\,ppm ($\log X_{\mathrm{N_2 O}} = -4.64^{+1.54}_{-4.24}$). We note that our Visit 1 retrieval finds no evidence of HCN, which was previously suggested from the \hst WFC3 analysis of \planetname by \citet{Swain2021}. However, our G395H data does not rule out a scattering slope below 2\,$\micron$ (as shown by the wide 2$\sigma$ confidence region in Figure~\ref{fig:retrievals}) similar to that inferred by \citet{Swain2021} (but cf. \citealt{LibbyRoberts2022} and \citealt{Mugani2021}).

In contrast, our Visit 2 retrieval (right panels of Figure~\ref{fig:retrievals}) is consistent with a flat line with no constraints on the atmospheric composition. Such a flat spectrum can be explained by many degenerate atmospheric properties, including a high mean molecular weight, low temperature, low surface pressure, and/or a high-altitude aerosol layer. Given that our data are sufficiently precise to differentiate between several high mean molecular weight atmospheres with high surface pressure / deep clouds (as shown by our inference of a H$_2$O-dominated atmosphere from Visit 1), our favored explanation for the featureless Visit 2 spectrum is a high-altitude aerosol layer. However, a wide range of cloud top pressures are permitted ($\log P_{\rm{cloud}} < -1.3$ to 1$\sigma$; see Figure~\ref{fig:retrievals}) after marginalization over all the possible combinations of atmospheric temperature and background gases with higher mean molecular weight than H$_2$O.

Assuming \planetname's transmission spectra are explained by a planetary atmosphere, our retrievals thus suggest that the cloud opacity would need to significantly increase between Visit 1 and 2 to explain our different spectra. As further discussed in Section~\ref{ssec:planetvar}, it is highly improbable that an atmosphere can change from a relatively clear state in one visit to host a global high-altitude cloud in the next visit. Furthermore, \planetname's equilibrium temperature places it in a parameter space without obvious condensable material for clouds to form \citep[e.g.,][]{gao2021}, which would suggest a photochemical haze as the cause of this aerosol layer. Similarly, a transition from a relatively clear atmosphere to one with such high haze opacity would be highly improbable given no change in radiative forcing. However, given the wide uncertainty on our retrieved cloud-top pressure, we note that both Visits 1 and 2 are consistent with an intermediate cloud pressure at $\sim$ 10\,mbar to 1$\sigma$. We next turn to consider an alternative explanation that does not require an atmosphere.

\subsubsection{Unocculted Starspot Scenario} \label{ssec:retrievals_spot}

We next investigate the alternative explanation of unocculted starspots shaping the observed transmission spectra, adopting the same retrieval configuration as \citetalias{Moran2023} (\citeyear{Moran2023}). This retrieval model assumes no planetary atmosphere, with unocculted stellar heterogeneities producing any wavelength-dependent features in the transmission spectrum \citep[see][for a review of the impact of starspots on transmission spectra]{Rackham2023}. Four parameters define this model (priors in brackets): the heterogeneity temperature, $T_{\rm{het}}$ ($\mathcal{U}$(2300\,K, 1.2\,$T_{*, \rm{eff}}$)), the heterogeneity covering fraction, $f_{\rm{het}}$ ($\mathcal{U}$(0.0, 0.6)), the photospheric temperature, $T_{\rm{phot}}$ ($\mathcal{N}$($T_{*, \rm{eff}}$, $\sigma_{T_{*, \rm{eff}}}$)), and the planetary radius, $R_p$ ($\mathcal{U}$(0.9\,$R_{\rm{p, \, obs}}$, 1.1\,$R_{\rm{p, \, obs}}$)). The priors are specified in terms of literature properties of the host star: $T_{*, \rm{eff}} = 3270$\,K and $\sigma_{T_{*, \rm{eff}}} = 140$\,K \citep{Bonfils2018}. We calculate the impact of the transit light source effect \citep{Rackham2018} by interpolating PHOENIX models \citep{Husser2013} using the \texttt{PyMSG} package \citep{Townsend2023}. We verified that a more complex parameterization of stellar contamination, including both faculae and spots, does not improve the fit or alter the retrieved spot properties, so we focus here on results assuming a single-heterogeneity population. As in the previous section, we perform independent retrievals for the two visits using the \eureka reduction and without a free offset between the detectors. 

Our Visit 1 retrieval is well-explained by unocculted starspots covering $\sim$ 20\% of the stellar surface with a spot temperature $\sim$400\,K cooler than the photosphere. These starspot properties are consistent with the stellar spectrum fits described in Section~\ref{ssec:stellar}. As shown in Figure~\ref{fig:retrievals}, the starspot scenario explains the Visit 1 data with a spectral slope shortwards of 4\,$\micron$. The posterior distributions in Figure~\ref{fig:retrievals} demonstrate that a wide range of spot coverage fractions are consistent with Visit 1 ($f_{\rm{het}} = 0.23^{+0.21}_{-0.10}$) --- due to the $f_{\rm{het}}$--$T_{\rm{het}}$ degeneracy \citep[e.g.][their Figure 10]{Rathcke2021}.

However, once again our Visit 2 retrieval is consistent with a flat, featureless spectrum. Under the starspot model, this requires either a low spot coverage fraction or a spot temperature similar to the stellar photosphere. However, the retrieved starspot fraction for Visit 2 is formally consistent with Visit 1 within 1\,$\sigma$, which agrees with the out-of-transit stellar modeling in Section~\ref{ssec:stellar} that found a negligible difference between visits.

\subsubsection{An Atmosphere vs. Starspots} \label{ssec:retrievals_fit_comparison}

At first glance, the available statistical evidence equally supports the atmosphere scenario or the starspot scenario for Visit 1. The Bayesian evidences ($\ln \mathcal{Z} = 490.1$ and $490.3$ for the atmosphere vs. starspot models, respectively) and minimum reduced $\chi^2$ ($\chi^2_{\nu} = 1.35$ and $1.36$ for the atmosphere vs. starspot models, respectively) are indistinguishable. However, this is largely a manifestation of the higher dimensionality of our reference atmosphere model compared to the starspot model (15 parameters vs. 4 parameters for the starspot model). Comparing the $\chi^2$ directly, we find the maximum likelihood atmospheric scenario model ($\chi^2 = 61$ with 45 dof) achieves a better quality fit than the starspot scenario ($\chi^2 = 76$ with 56 dof). This better fit arises from the spectral slope caused by H$_2$O absorption in combination with the CH$_4$ feature near 3.3\,$\micron$ providing a better fit than the slope produced by unocculted starspots. 

We, therefore, ran an additional, simplified, atmosphere scenario retrieval focusing only on those properties indicated by the data. Our motivation is to account for the `Occam penalty' disfavoring models with redundant free parameters (i.e., the non-detected molecules in the atmosphere scenario). We thus consider a H$_2$O-dominated atmosphere with trace CH$_4$ and N$_2$O, constituting a 4-parameter model defined by the planetary reference radius, temperature, and the CH$_4$ and N$_2$O abundances (the H$_2$O abundance determined via abundances summation). This simplified model obtains an excellent fit to Visit 1 ($\chi^2 = 63$ with 56 dof; $\chi^2_{\nu} = 1.13$; $\ln \mathcal{Z} = 494.2$), providing tentative evidence favoring an atmosphere over starspots for Visit 1. The lack of observed starspot variability in the out-of-transit stellar spectrum (Figure~\ref{fig:phoenix}), in small tension with the low spot fraction required to render Visit 2 flat (Figure~\ref{fig:retrievals}), may also support the atmosphere interpretation. 

However, the inferred atmosphere from Visit 1 --- a H$_2$O-dominated atmosphere with trace CH$_4$ and N$_2$O --- would be unstable against a runaway greenhouse, as all three molecules are highly effective greenhouse gases and highly susceptible to photolysis \cite[e.g.,][]{Rugheimer2015}. Thus, such an atmosphere would rapidly be lost to space \citep[e.g.,][]{Goldblatt2013}, requiring ongoing outgassing, a very high ($>$5 wt\%) initial H$_2$O inventory at formation, and a present-day magma ocean \citep{Schaefer2016} to replenish the H$_2$O to the levels suggested by our Visit 1 atmospheric retrieval. Moreover, an H$_2$O-dominated atmosphere would be expected to also include ample oxidized carbon species, such as CO$_2$ or CO, given outgassed abundances suggested from previous studies \citep{Sossi2023,Tian2023}. While we do not see evidence for CO$_2$, our NIRSpec G395H data cannot constrain the presence or absence of CO. The co-presence of $\sim$ percent levels of CH$_4$ with CO in an H$_2$O-dominated atmosphere would require finely tuned carbon abundances and oxygen fugacities of the planetary interior \citep{Tian2023}.

Given the disagreement between our two visits, we suggest it is premature to claim a clear preference for an atmosphere or starspots. Nevertheless, our JWST observations do rule out the H$_2$-dominated atmosphere with HCN previously suggested by \cite{Swain2021} (see Appendix~\ref{app:sec:previousdata}). One path to a resolution is the significant predicted deviations between the atmosphere and starspot scenarios at shorter wavelengths (see Figure~\ref{fig:retrievals}), which could be probed with future observations.

\subsection{Potential For Planetary Variability} \label{ssec:planetvar}

The impact of atmospheric variability on transmission spectra of tidally locked terrestrial planets has been explored by numerous teams \citep[e.g.][]{May2021, Song2021, Cohen2022, Rotman2023}, with the general conclusion being that such variability is below the detectability limit of \jwst instruments. This is because, while the cloud cover of any one region of the terminator can be highly variable, we are observing limb averaged spectra that simultaneously probe cloudy and cloud-free regions. With the difference between \planetname's two spectra of order $\sim$100\,ppm at the 4.5\,$\micron$ ``feature'' (see Figure \ref{fig:tspec}), any planetary variability would be required to be at least an order of magnitude larger than predicted by current GCMs of temperate, tidally locked terrestrial planets in the above works. While \planetname is hotter than the models in the above works, it remains physically unlikely that such variability is the cause of the observed spectral differences. This is because at the equilibrium temperature of \planetname the planetary dayside is more likely to be cloud free. Further, such large scale variability in transmission has yet to be detected on even the most optimal targets -- hot Jupiters \citep[e.g.][]{Kilpatrick2020}. 

\section{Conclusions} \label{sec:conclusions}

We presented two transit observations of the super-Earth \planetname with \jwst NIRSpec G395H that yield distinctly different transmission spectra. Barring other possibilities, the differences between the visits may be explained by random noise fluctuations that took the unfortunate shape of spectral features for one visit. Without a third observation of \planetname at these wavelengths, it is impossible to determine which of the two visits most accurately reflects the true nature of the planet. Our conflicting observations demonstrate the potential risk of claiming atmospheric detections for rocky exoplanets based on a single \jwst observation.

Should Visit 1 represent the ``truth'' for \planetname\, our retrievals exhibit a slight preference for a H$_2$O-dominated atmosphere with trace CH$_4$ and N$_2$O ($\chi^2_{\nu} = 1.13$) compared to contamination from unocculted starspots ($\chi^2_{\nu} = 1.36$). These two scenarios would produce significantly different transmission spectra at shorter wavelengths (see Figure~\ref{fig:retrievals}). While \hst WFC3 and ground-based optical data exist for \planetname, their lack of wavelength overlap with our NIRSpec G395H observations results in too much freedom when accounting for a possible inter-instrument transit depth offset, resulting in either scenario being allowed (see Appendix~\ref{app:sec:previousdata}). However, our Visit 1 observations do rule out a H$_2$-dominated atmosphere containing HCN, as previously suggested by the \hst WFC3 analysis of \cite{Swain2021}.  See Appendix \ref{app:sec:previousdata} for further comparison of our new \jwst observations to the existing \hst WFC3 and ground-based data.

Instrument systematics could provide an alternative explanation for our divergent transmission spectra. Allowing for a possible offset between the NRS1 and NRS2 detectors would eliminate the statistical significance of the bluewards slope in the Visit 1 data (see Section~\ref{ssec:datadiffs} and Appendix~\ref{app:sec:gaussians}), which drives our inference of an atmosphere or unocculted starspots. Larger detector offsets were seen between NRS1 and NRS2 in \citetalias{Moran2023} (\citeyear{Moran2023}), but were manually corrected based on differences between the visits and not explored in their equivalent Gaussian feature tests. However, we note that we do not see evidence during our data reduction process for the need for a similar superbias correction, and find equivalent spectra regardless of applying such a correction. Future observations of \planetname with NIRSpec G395M would cover the same wavelengths --- without the detector gap --- to determine the true shape of \planetname's transmission spectrum at these wavelengths. With the growing evidence for detector offsets within NIRSpec G395H data, it may be preferable to observe planets with small predicted spectral features using NIRSpec G395M instead, especially if the star is dim enough to allow it, to ensure that detector offsets are not mistaken for atmospheric features.  

Additional future observations at wavelengths below 3\,$\micron$, but still overlapping with the NIRSpec G395H wavelength range, are crucial to breaking the degeneracy between the atmospheric and unocculted starspot explanations. This atmosphere--starspot degeneracy has now been seen in multiple NIRSpec G395H observations of small planets orbiting M dwarfs (see also \citetalias{Moran2023}, \citeyear{Moran2023}). For example, NIRISS SOSS observations combined with our existing NIRSpec G395H data, or new NIRSpec G395M data (avoiding any potential detector offsets), will be crucial to determine if (1) \planetname's spectral features are real and repeatable, and (2) if these features are best described by a planetary atmosphere or stellar contamination. 

While the quest continues for an unambiguous atmospheric detection on a rocky planet, our results for \planetname provide an important reminder of the necessity of repeat observations to confirm the reliability of potential detections.
\section*{Acknowledgments}

This work is based in part on observations made with the NASA/ESA/CSA \jwst. The data were obtained from the Mikulski Archive for Space Telescopes at the Space Telescope Science Institute, which is operated by the Association of Universities for Research in Astronomy, Inc., under NASA contract NAS 5-03127 for JWST. This research has made use of the NASA Exoplanet Archive, which is operated by the California Institute of Technology, under contract with the National Aeronautics and Space Administration under the Exoplanet Exploration Program. Support for \jwst program \#1981 was provided by NASA through a grant from the Space Telescope Science Institute, which is operated by the Association of Universities for Research in Astronomy, Inc., under NASA contract NAS 5-03127. We thank the anonymous referee for their constructive and timely feedback, alongside fruitful discussions that consequently improved our study. R.J.M. is supported by NASA through the NASA Hubble Fellowship grant HST-HF2-51513.001, also awarded by the Space Telescope Science Institute, which is operated by the Association of Universities for Research in Astronomy, Inc., for NASA, under contract NAS 5-26555. Work by S.P. is supported by NASA under award number 80GSFC21M0002. J.K. acknowledges financial support from Imperial College London through an Imperial College Research Fellowship grant. This material is based upon work performed as part of NASA’s CHAMPs team, supported by the National Aeronautics and Space Administration (NASA) under Grant No. 80NSSC21K0905 issued through the Interdisciplinary Consortia for Astrobiology Research (ICAR) program. The authors thank Raissa Estrela and the Excalibur team at JPL for useful discussions on the HST model presented in \citet{Swain2021}.
\facilities{JWST (NIRSpec)}

\vspace{0.3cm}
\noindent The \jwst data presented in this paper were obtained from the Mikulski Archive for Space Telescopes (MAST) at the Space Telescope Science Institute. The specific observations analyzed can be accessed via \dataset[DOI: 10.17909/0njr-8110]{https://doi.org/10.17909/0njr-8110}.

\software{\\ Astropy \citep{astropy,astropy2},
\\ \texttt{batman} \citep{Kreidberg2015},
\\ \texttt{CHIMERA} \citep{line2013,line2014},
\\ \texttt{Dynesty} \citep{Speagle2020},
\\ \texttt{emcee} \citep{emcee2013},
\\ \eureka \citep{Bell2022},
\\ ExoCTK \citep{exoctk},
\\ \exotic \citep{Alderson2022},
\\ \texttt{ExoTiC-LD} \citep{Grant2022},
\\ \firefly \citep{Rustamkulov2022}
\\ IPython \citep{ipython},
\\ \texttt{jwst} \citep{JWST_pipeline},
\\ \texttt{lacosmic} \citep{lacosmic},
\\ \texttt{lmfit} \citep{Newville2014}, 
\\ Matplotlib \citep{matplotlib},
\\ NumPy \citep{numpy, numpynew},
\\ \texttt{PHOENIX} \citep{Allard2012},
\\ \texttt{PICASO} \citep{Batalha2019},
\\ \texttt{POSEIDON} \citep{MacDonald2017, MacDonald2023},
\\ \texttt{PyMultiNest} \citep{Feroz2009,Buchner2014},
\\ SciPy \citep{scipy}
}

%%%%%%%%%%%%%%%%%%%%%%%%%%%%%%%%%%%%%%%%
\bibliography{GJ1132b}{}
\bibliographystyle{aasjournal}
%%%%%%%%%%%%%%%%%%%%%%%%%%%%%%%%%%%%%%%%

%%%%%%%%%%%%%%%%%%%%%%%%%%%%%%%%%%%%%%%%%%%%%%%%
%%%%%%%%%%%%%%%%    APPENDIX    %%%%%%%%%%%%%%%%
%%%%%%%%%%%%%%%%%%%%%%%%%%%%%%%%%%%%%%%%%%%%%%%%
\newpage
\appendix

\section{Data Analysis}
\subsection{Light Curve Fitting: To Spectrally Bin Before or After?} \label{sec:binning}
\begin{figure*}[!htb]
    \centering
    \includegraphics[width = 0.95\textwidth]{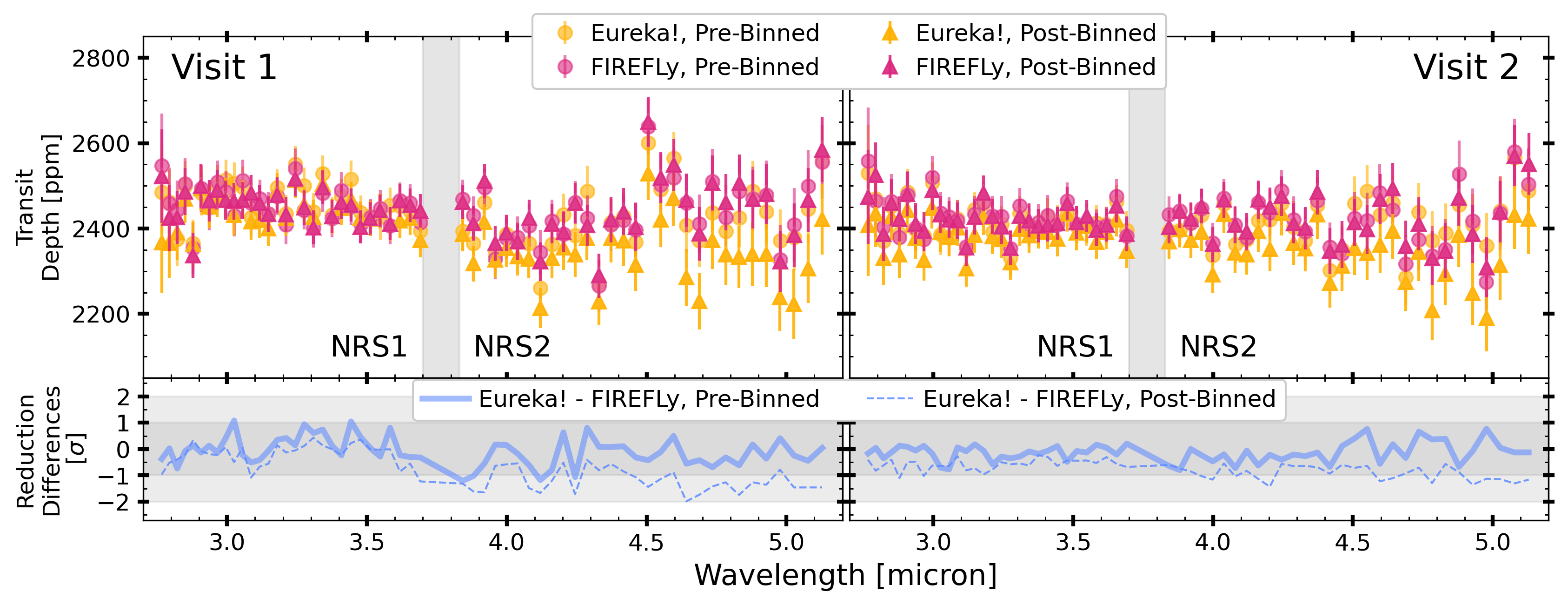}
    \caption{\textbf{Comparison of \eureka and \firefly transmission spectra from binning before and after light curve fitting.} In yellow are the \eureka reductions and in pink are the \firefly reductions. For both, the shaded circles are the reductions that are first binned to R$\sim$ 100 and then fit (pre-binned) and the darker triangles are the reductions that are fit at the native resolution and then binned to R$\sim$100 (post-binned). In the bottom panels we show the differences between the \eureka and \firefly reductions for both cases in units of $\sigma$, with the solid thicker line the pre-binned case and the thinner dashed line the post-binned case. The shaded regions denote $\pm$1$\sigma$ and $\pm$2$\sigma$ to guide the eyes. The \eureka and \firefly pre-binned spectra are clearly more consistent with each other (always within a $\pm$1$\sigma$ difference) than the post-binned case, suggesting that fitting at the native resolution is not always appropriate.}
    \label{app:fig:binning}
\end{figure*}

Light curves from \eureka and \firefly were extracted at both the R$\sim$100 resolution and at the native pixel level resolution. The native pixel level resolution light curves were fit in the same manner as described in Sections \ref{ssec:eureka} and \ref{ssec:firefly} and then binned to the same wavelength grid as the R$\sim$100 spectra to compare the effect of binning before and after light curve fitting. Figure \ref{app:fig:binning} shows the differences between \eureka and \firefly transmission spectra with these two methods. We find that the transmission spectra diverge significantly on the red side of the NRS2 detector when fitting at the native pixel resolution and then spectrally binning, likely due to the lower SNR at longer wavelengths. This suggests that fitting at the native resolution is not always appropriate.

Notably, it is the \eureka reduction that changes the most when binning before or after light curve fitting.  We conducted a further comparison using a least squared fitting routing in \eureka (instead of \texttt{emcee}) and find better agreement with \firefly in the native pixel light curve fits. However, an analysis of the \texttt{emcee} chains for both resolutions shows no evidence that either fit has not converged and no multi-modal distributions are present in the posteriors. Nevertheless, because the resulting spectra were consistent when binning prior to light curve fitting regardless of the tool used to perform the light curve fitting, we chose to adopt the pre-binned R$\sim$100 spectra in our transmission spectra analysis.

Choices in limb darkening treatment (fixing vs. fitting) have been shown to impact \jwst transmission spectra (e.g. Carter \& May et al., in prep.) for larger planets. Here we apply their recommendations, specifically to fix limb darkening rather than fit for it, and to pre-bin the data to a lower resolution. This minimizes additional uncertainty and increases accuracy in the final transmission spectrum. We do not see significant differences between the pre-binned \firefly and \eureka spectra even with the different treatment of limb darkening between the two reductions (\firefly uses the white light parameters for all spectroscopic bins, \eureka fixes to wavelength dependent limb darkening based on stellar models).  It is therefore unlikely that our limb darkening choices are causing the differences between the pre- and post-binned transmission spectra in this work.

\pagebreak
\subsection{2D Light Curve Comparisons} \label{app:sec:2d}
\begin{figure*}[!htb]
    \centering
    \includegraphics[width = \textwidth]{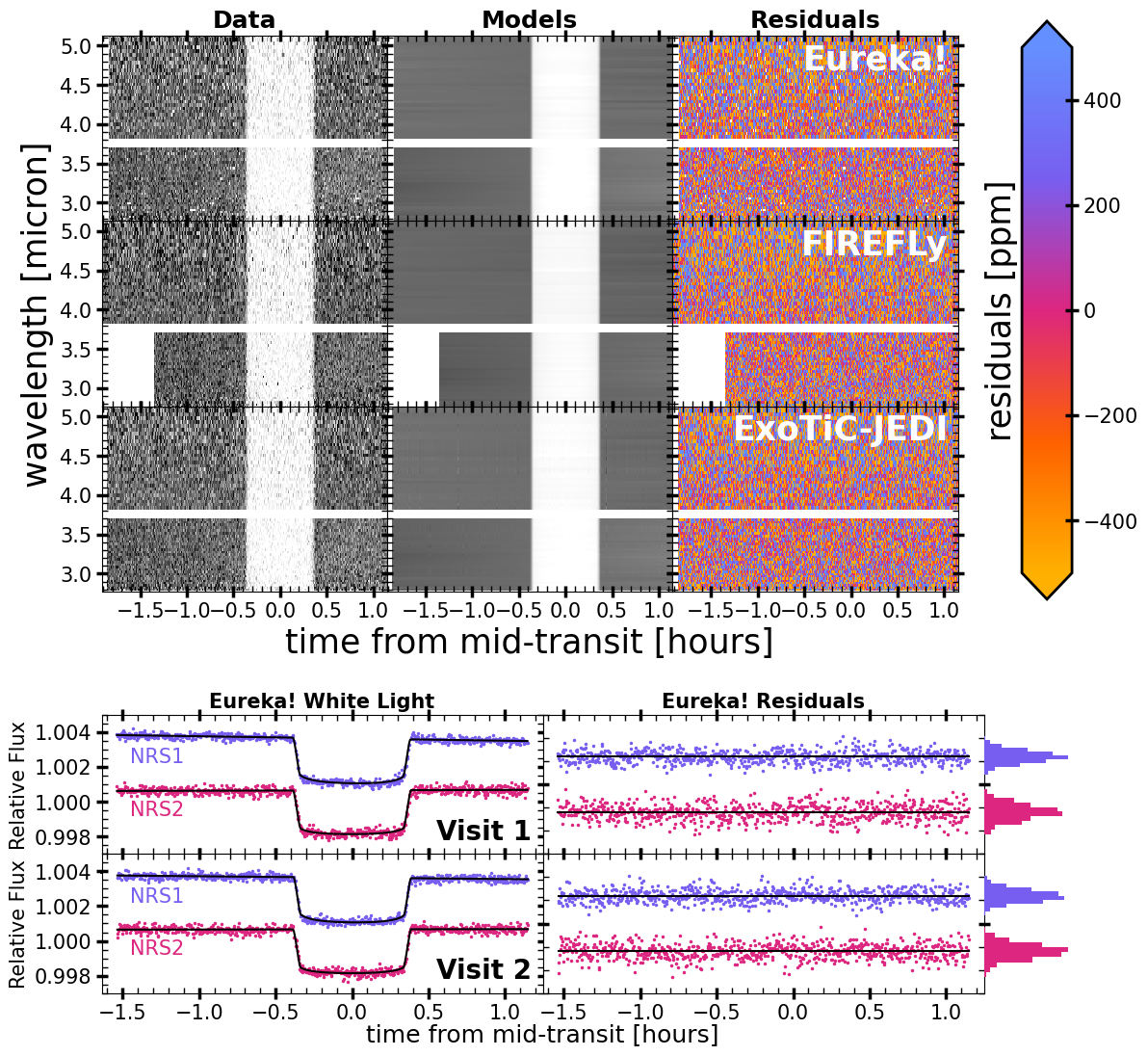}
    \caption{\textbf{2D light curves for all reductions from Visit 1 and \eureka white light curves for both visits}. In the top panel we show the 2D light curves for all three reductions, including the data (left), models (middle) and residuals (right) at the R$\sim$100 resolution. \firefly trims extra data from the start of the NRS1 data due to a temporal ramp. In the bottom panel we show the \eureka white light curves and models for NRS1 and NRS2 (left), the residuals (middle), and a histogram of the residuals for both NRS1 and NRS2 (right) for both visits 1 and 2. There is no evidence of occulted starspots in any of the white light curves.}
    \label{app:fig:2d}
\end{figure*}

Figure \ref{app:fig:2d} shows the R$\sim$100 2D light curves for all three reductions for Visit 1. We note that while each analysis has different levels of correlated noise in the data panel (left), the noise is generally well modeled as seen by the models and residuals panels (middle and right). In the bottom panel we show the \eureka white light curves from Visits 1 and 2 to highlight the precision of the data. The residuals (lower middle) are well described by Gaussians (lower right).

\pagebreak

\subsection{Best-Fit Orbital Parameters} \label{app:sec:bestfits}

In Table~\ref{tab:orbital_parameters} we include the best-fit orbital parameters from \eureka, \firefly, and \exotic. \eureka performs a joint fit across the two visits independently for the two detectors (a total of two white light fits) while \firefly and \exotic fit all four white light curves independently. The values provided in Table~\ref{tab:orbital_parameters} are weighted means of the respective number of white light curve fits from each pipeline.

For spectroscopic light curve fitting, \eureka and \firefly use the values in Table~\ref{tab:orbital_parameters} for both visits and detectors while \exotic uses the best-fit white light parameters from the respective white light fit for each visit and detector. 

We note that because the scaled semi-major axis (a/R$_s$) and orbital inclination (i) are by definition degenerate with one another, all three pipelines converge to a different part of parameter space. Regardless of these differences in a/R$_s$ and inclination, no transit depth offset is required to align the spectra from the three reductions (see Figure \ref{fig:tspec}).

\begin{deluxetable}{rlrlrlrl}
\tablewidth{0pt}
\tablecaption{Best fit orbital parameters from white light curve fitting. \label{tab:orbital_parameters}}
\tablehead{
    \multicolumn{2}{c}{Parameter} & 
    \multicolumn{2}{c}{\eureka} & 
    \multicolumn{2}{c}{\firefly} & 
    \multicolumn{2}{c}{\exotic}
}
\startdata
R$_p$/R$_s$ & [unitless] & 0.04917 & $\pm$ 8.79$\times$10$^{-5}$ & 0.04935 & $\pm$ 2.72$\times$10$^{-3}$ & 0.04945 & $\pm$ 1.2$\times$10$^{-4}$ \\
(R$_p$/R$_s$)$^2$ & [ppm] & 2418 & $\pm$ 9 & 2435 & $\pm$ 268 & 2445 & $\pm$ 12 \\
T$_0$ - 2460000 & [BJD$_{TDB}$] & 0.97650 & $\pm$ 1.38$\times$10$^{-5}$ & 0.976537 & $\pm$ 2.0$\times$10$^{-5}$ & 0.976598 & $\pm$ 2.59$\times$10$^{-5}$ \\
& & -- & & 9.121152 & $\pm$ 1.9$\times$10$^{-5}$ & 9.121220386 & $\pm$ 2.50$\times$10$^{-5}$ \\
Period & [days] & 1.628931 & (fixed) & 1.628931 & (fixed) & 1.628931 & (fixed) \\
a/R$_s$ & [unitless] & 15.04 & $\pm$ 0.21 & 15.62 & $\pm$ 0.29 & 14.70 & $\pm$ 0.37 \\
i & [degrees] & 88.14 & $\pm$ 0.11 & 88.39 & $\pm$ 0.14 & 87.97 & $\pm$ 0.21 \\
\enddata
\tablecomments{We adopt the \eureka results as our system parameters. The \eureka best-fit values are derived from a mean of independent joint-visit fits to the two detectors. Because \eureka performs a joint fit of the two visits, only one T$_0$ value is reported. The \firefly and \exotic values are derived from a weighted mean of four independent white light curve fits, one per detector for each visit.}
\vspace{-1cm}
\end{deluxetable}

\subsection{Correlated Noise} \label{app:sec:allan}

\begin{figure*}[!htb]
    \centering
    \includegraphics[trim = 0.0cm 0.5cm 0.0cm 0.2cm, width = 0.46\textwidth]{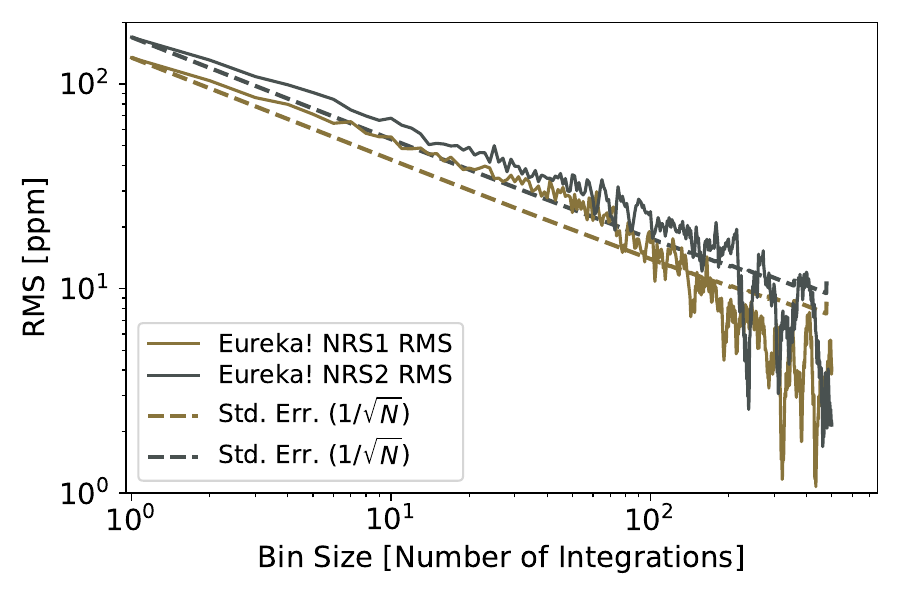}
    \includegraphics[trim = 0.0cm 0.5cm 0.0cm 0.2cm, width = 0.46\textwidth]{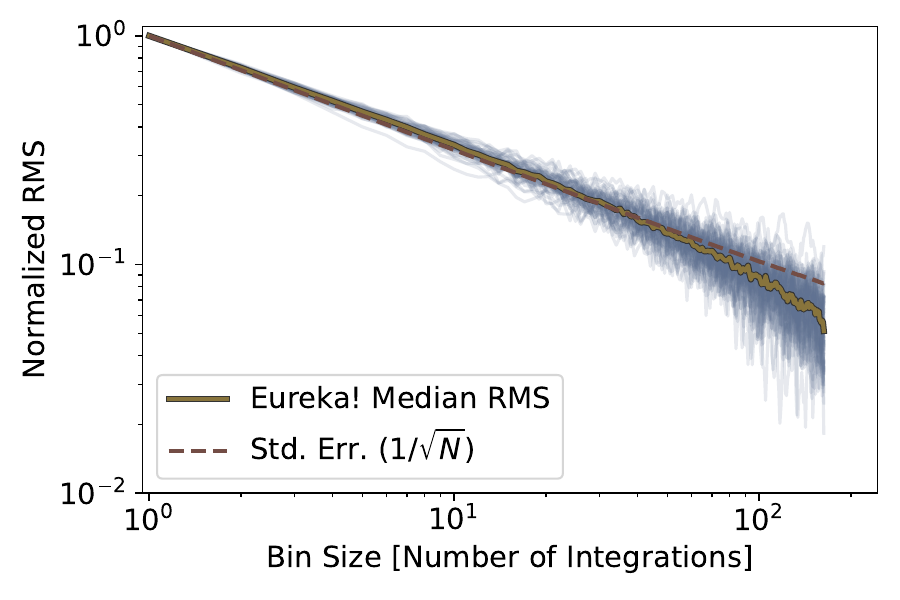}
    \caption{\textbf{Allan variance plots for the {\eureka} reduction.} 
    Left panel: variation of the NRS1 and NRS2 detector white light curve precision (root mean square; RMS) with longer bin timescales (more integrations). As seen in previous works (e.g., \citetalias{Moran2023} \citeyear{Moran2023}), correlated noise emerges around timescales of 0.5--5 minutes (5--50 integrations) and is likely due to thermal cycling \citep{Rigby2022}.
    Right panel: same, for spectroscopic light curves. The individual spectroscopic light curves show no evidence of correlated noise.}
    \label{app:fig:allan}
\end{figure*}

All three data reduction pipelines consider correlated noise by inflating error bars, if needed, at the white and spectroscopic light curve stage based on the scatter of the data. The Allan variance plot for \eureka's white and spectroscopic light curves (Figure~\ref{app:fig:allan}) demonstrates that the correlated noise is small enough to not impact our results. For \eureka, correlated noise in the white light curve appears when binning by timescales greater than 0.5--5 minutes, and this is believed to be due to some inherent thermal cycling from the telescope \citep{Rigby2022}. Because this timescale is much shorter than the transit duration, it is unlikely to affect the white light curve transit depth. Furthermore, correlated noise is negligible in the spectroscopic light curves.

\pagebreak
\section{Is It Flat? Gaussian Features and Detector Offsets} \label{app:sec:gaussians}

\begin{figure*}[!htb]
    \centering
    \includegraphics[width = \textwidth]{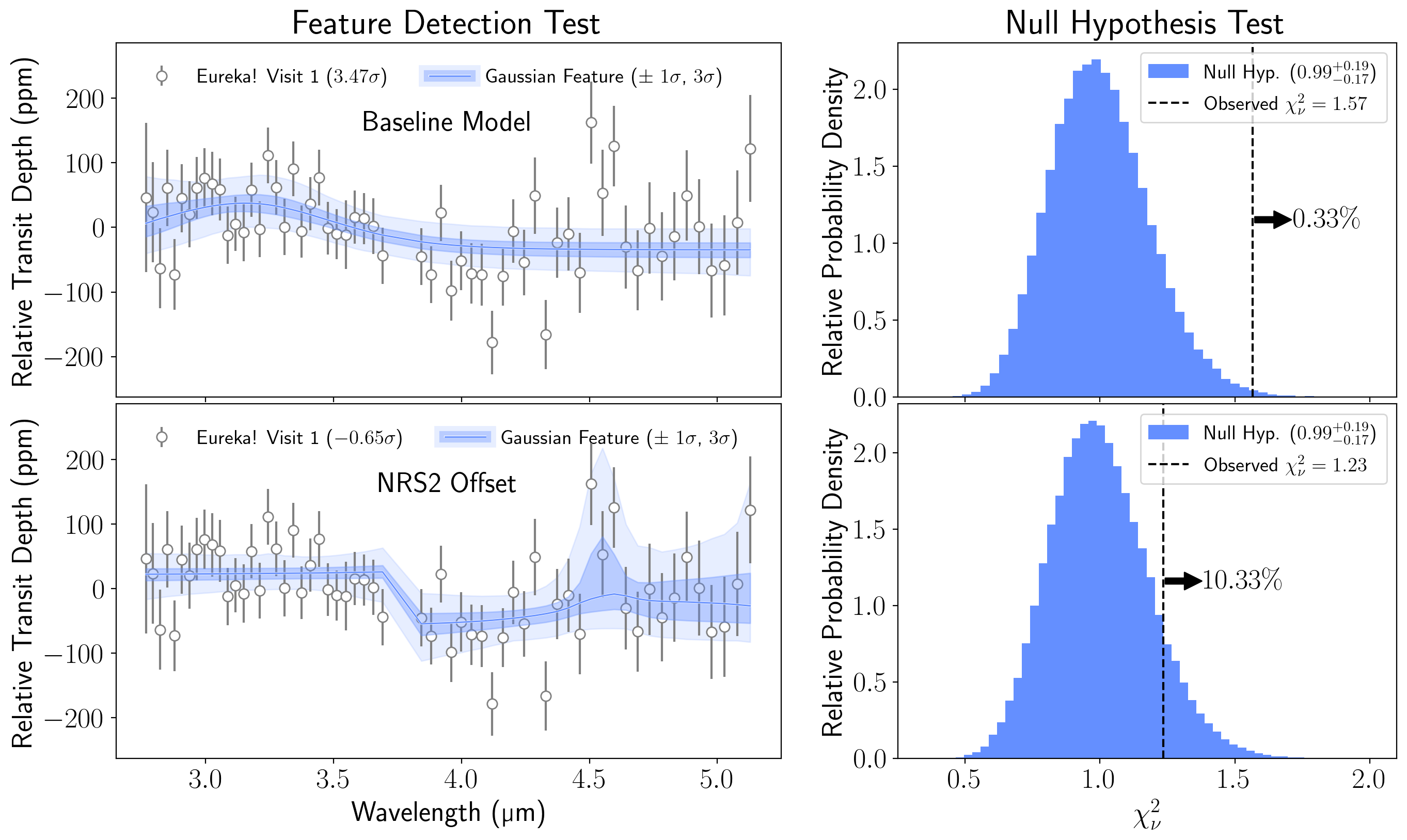}
    \caption{\textbf{Spectral feature detection tests for the \eureka Visit 1}. The top two panels show baseline results directly fitting the spectrum as shown, while the bottom two panels show results allowing a model offset for the NRS2 detector data ($>3.75$\,\micron). The left panels show the spectrum compared to the median best-fitting Gaussian feature model with 1$\sigma$ and 3$\sigma$ credibility envelopes displayed. Positive (negative) ``sigma'' values in the legend indicate the confidence with which the Gaussian feature model is favored (disfavored) compared to a featureless model (the null hypothesis). The right panels show the distribution of expected $\chi^2_{\nu}$ values assuming the null hypothesis. The observed $\chi^2_{\nu}$ is shown as the vertical dashed line along with the percent of the distribution that would be expected to have at least as large of a result. The complete figure set (9 images, for two visits plus the combined spectrum for three independent reductions) is available in the online journal.}
    \label{app:fig:flat}
\end{figure*}

Figure \ref{app:fig:flat} shows an example of the results presented in Section \ref{ssec:datadiffs} and Table \ref{tab:flat} that are used to determine the statistical significance of any candidate features in the transmission spectrum. Specifically, Figure \ref{app:fig:flat} highlights the \eureka reduction for our \planetname Visit 1 observations. The top row shows the nominal model (a single Gaussian feature), while the bottom row shows the model that allows for an offset between the NRS1 and NRS2 detectors (parameterized by a step function in the model). For both top and bottom, the left panels show the data compared to the best-fitting model, while the right panels show the distribution of expected reduced chi-squared values given the uncertainty on the data points. 

For the \eureka Visit 1 reduction shown, this demonstrates how allowing for the offset between the NRS1 and NRS2 detectors removes the significance of the Gaussian feature detection, and also changes the location of the Gaussian feature detected (from the blue wards slope to the feature near 4.5 $\micron$). A complete figure set showing analogous results for both visits plus the combined spectrum for each of the three independent reductions is available in the online journal. 

\figsetstart
\figsetnum{1}
\figsettitle{Gaussian Feature Detection Tests}

\figsetgrpstart
\figsetgrpnum{1.1}
\figsetgrptitle{\eureka Visit 1}
\figsetplot{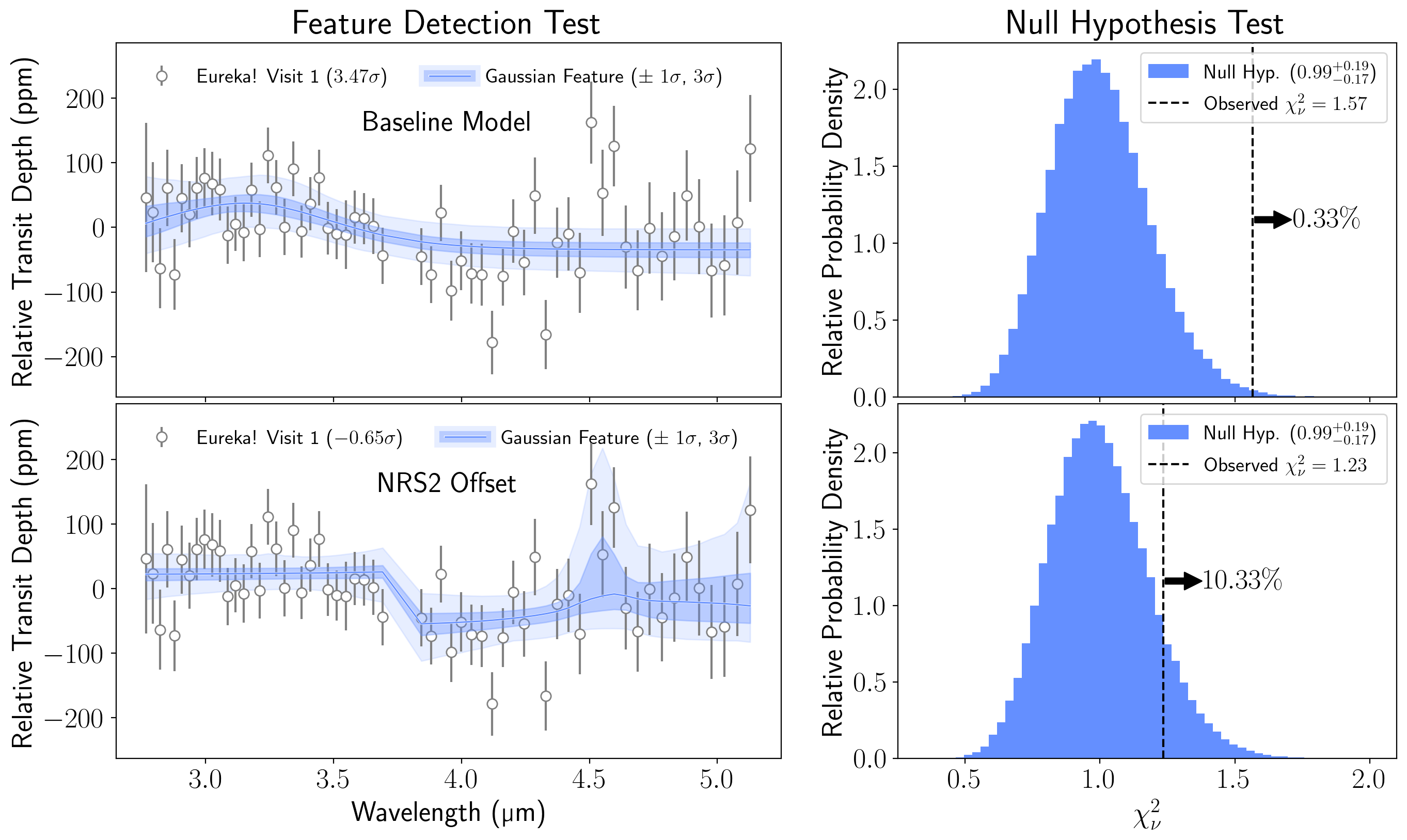}
\figsetgrpnote{\eureka Gaussian Feature Detection Tests for Visit 1 (as shown in Figure~\ref{app:fig:flat}).}
\figsetgrpend

\figsetgrpstart
\figsetgrpnum{1.2}
\figsetgrptitle{\eureka Visit 2}
\figsetplot{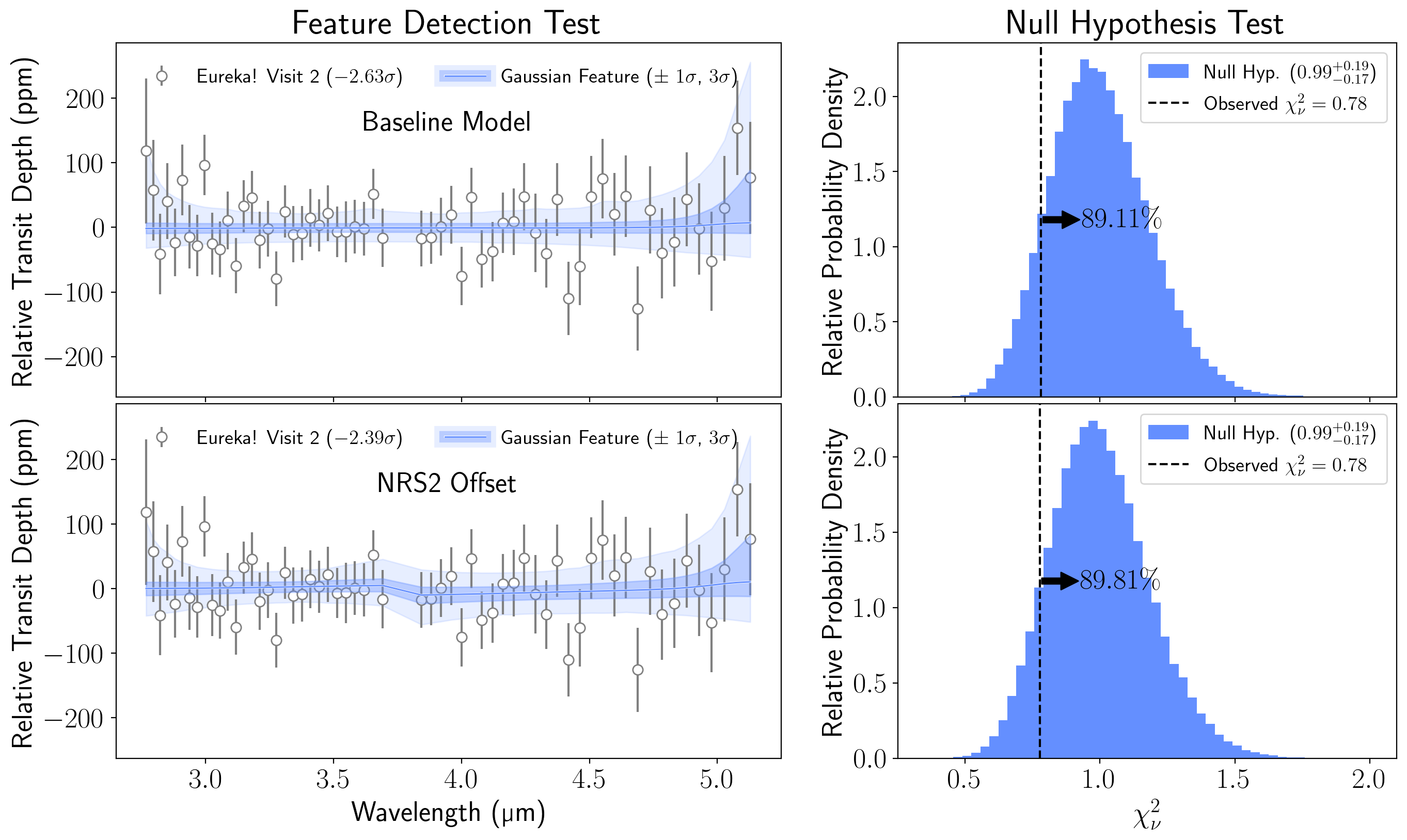}
\figsetgrpnote{\eureka Gaussian Feature Detection Tests for Visit 2 (as shown in Figure~\ref{app:fig:flat}).}
\figsetgrpend

\figsetgrpstart
\figsetgrpnum{1.3}
\figsetgrptitle{\eureka Combined Visit}
\figsetplot{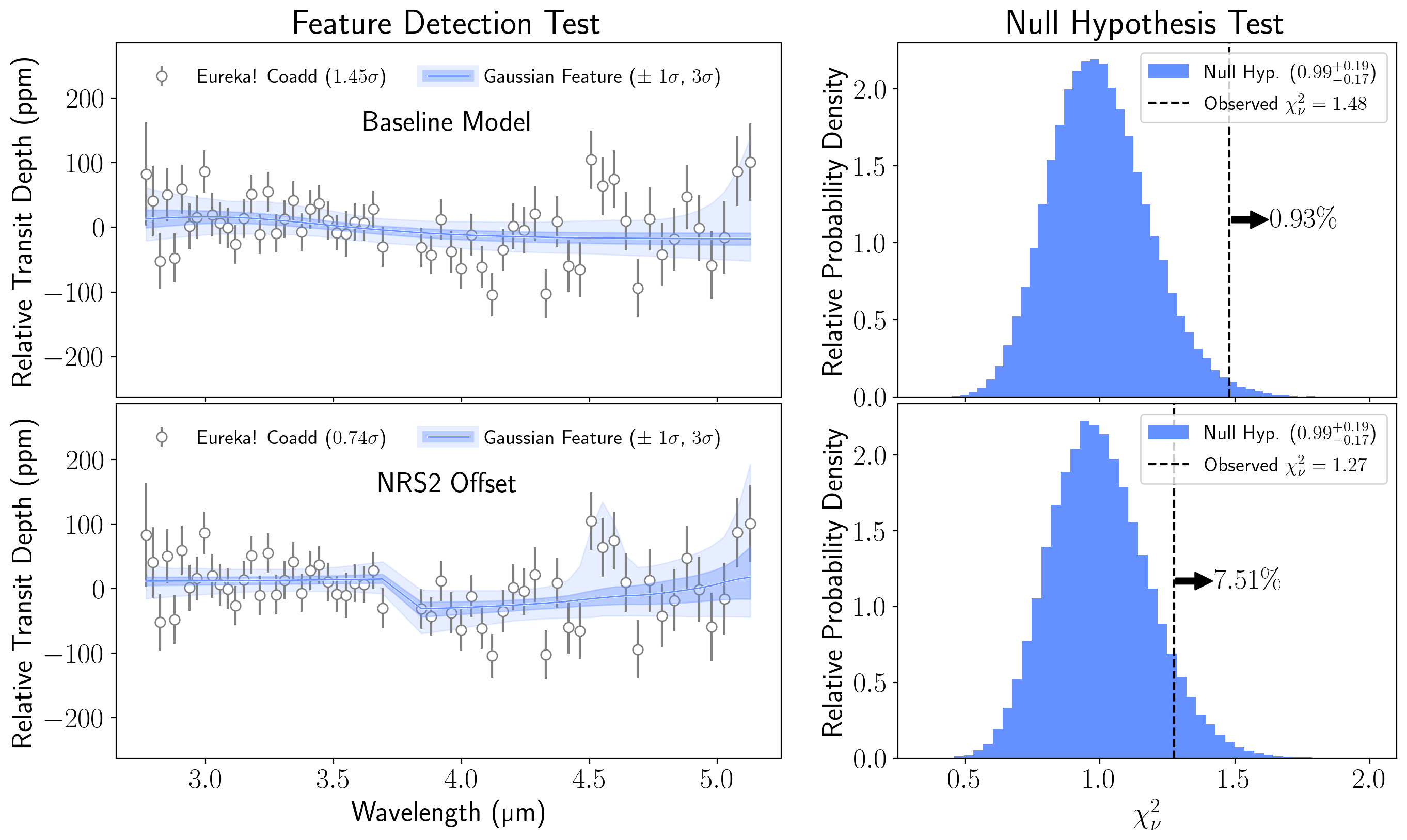}
\figsetgrpnote{\eureka Gaussian Feature Detection Tests for the Combined Visit (as shown in Figure~\ref{app:fig:flat}).}
\figsetgrpend

\figsetgrpstart
\figsetgrpnum{1.4}
\figsetgrptitle{\firefly Visit 1}
\figsetplot{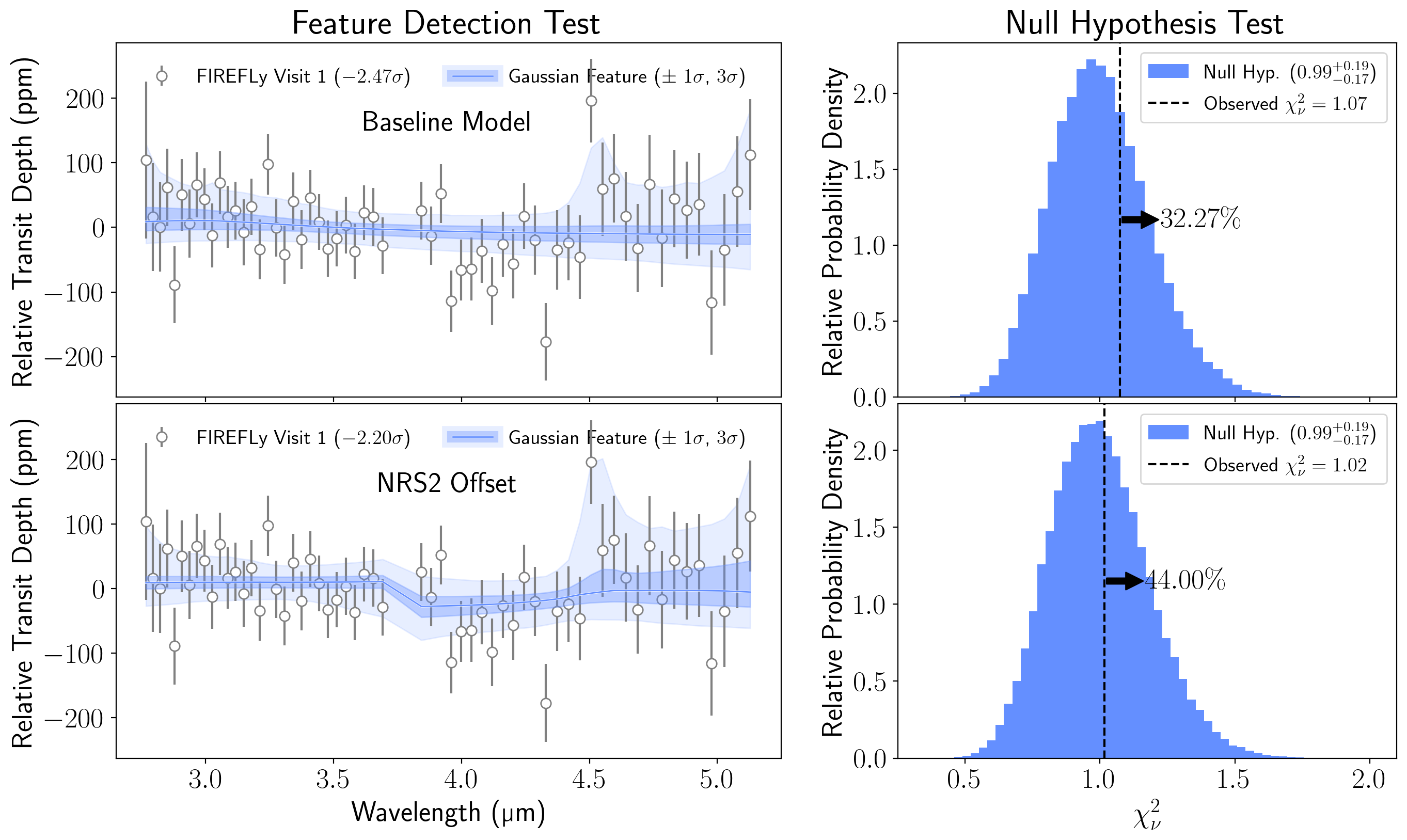}
\figsetgrpnote{\firefly Gaussian Feature Detection Tests for Visit 1 (as shown in Figure~\ref{app:fig:flat}).}
\figsetgrpend

\figsetgrpstart
\figsetgrpnum{1.5}
\figsetgrptitle{\firefly Visit 2}
\figsetplot{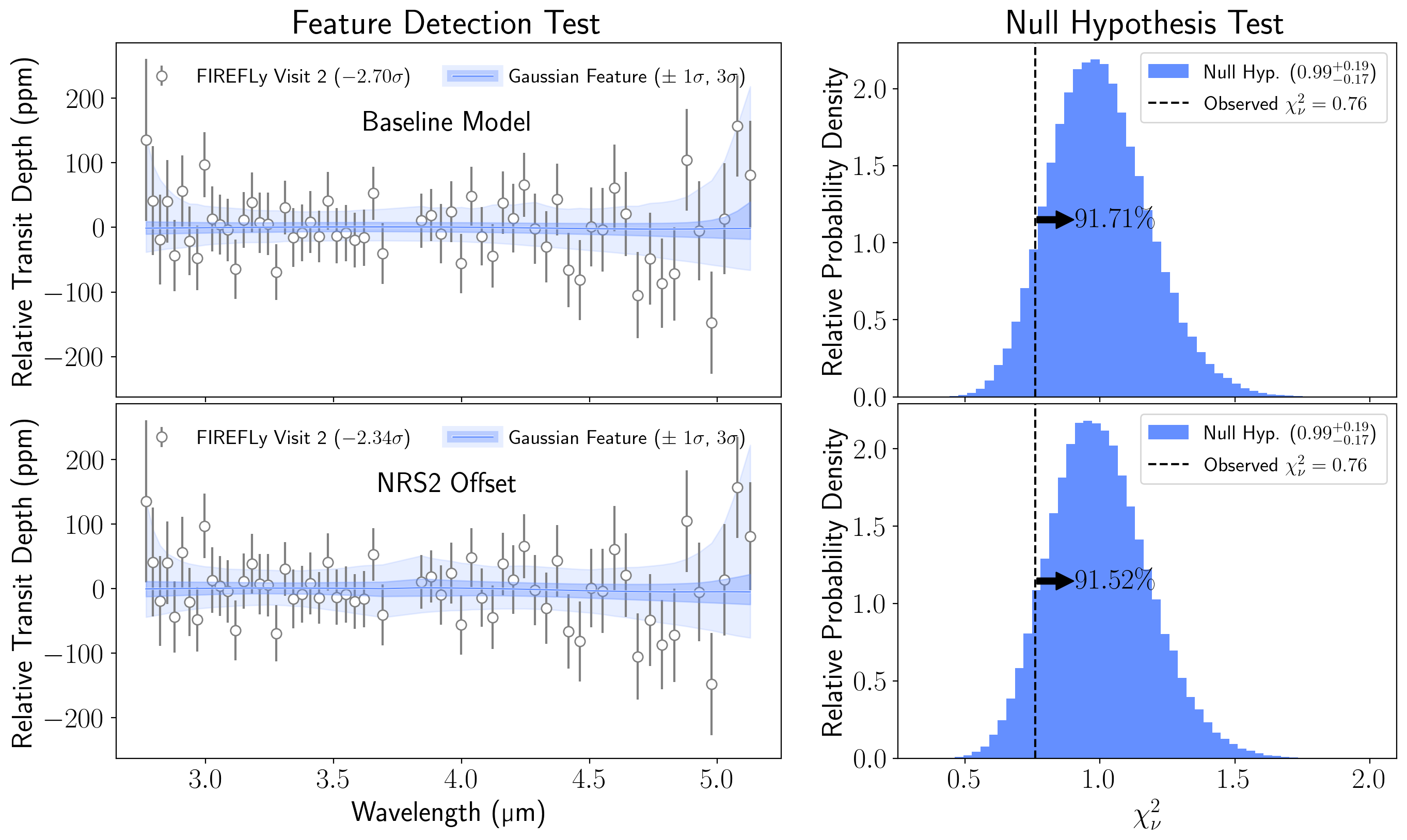}
\figsetgrpnote{\firefly Gaussian Feature Detection Tests for Visit 2 (as shown in Figure~\ref{app:fig:flat}).}
\figsetgrpend

\figsetgrpstart
\figsetgrpnum{1.6}
\figsetgrptitle{\firefly Combined Visit}
\figsetplot{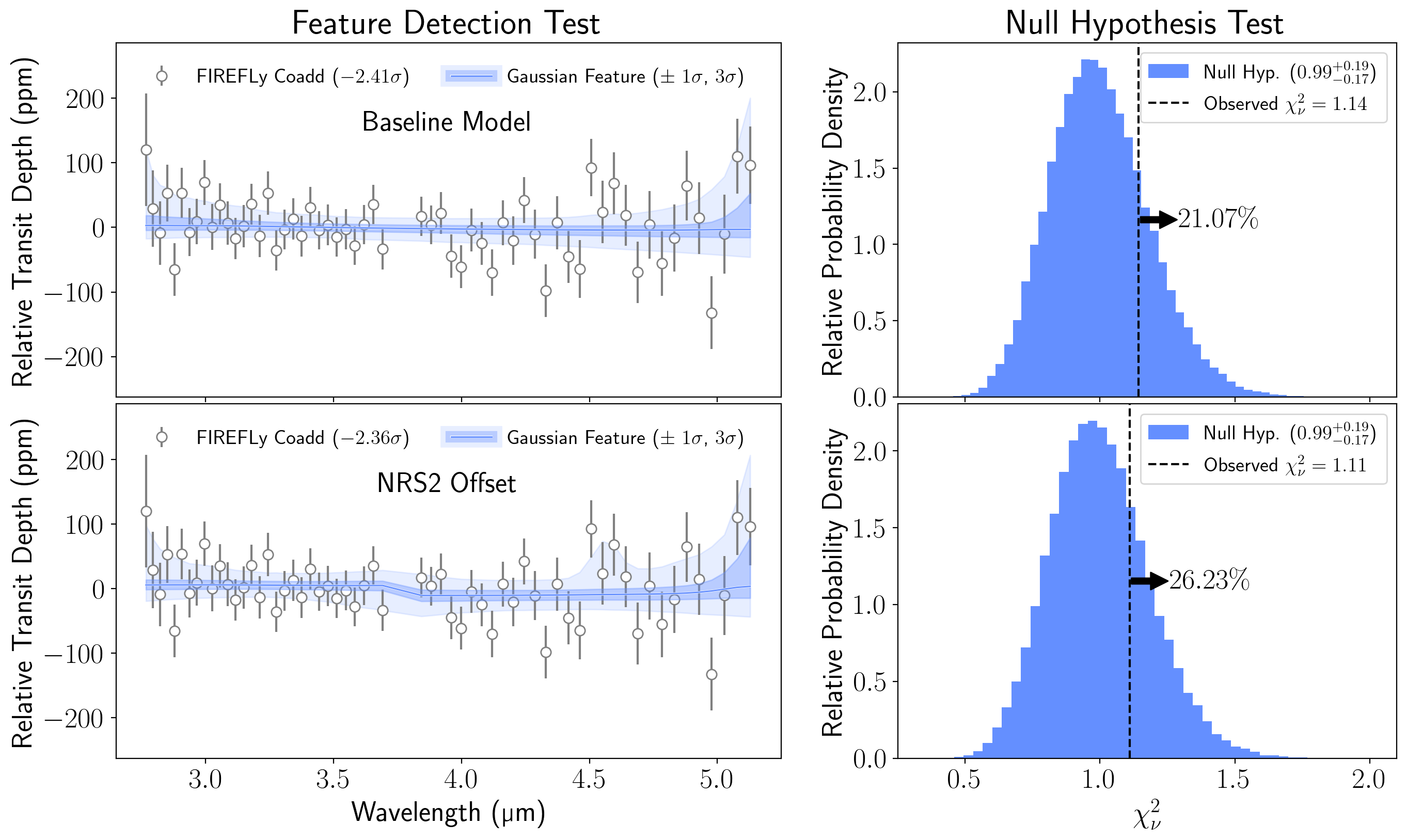}
\figsetgrpnote{\firefly Gaussian Feature Detection Tests for the Combined Visit (as shown in Figure~\ref{app:fig:flat}).}
\figsetgrpend

\figsetgrpstart
\figsetgrpnum{1.7}
\figsetgrptitle{\exotic Visit 1}
\figsetplot{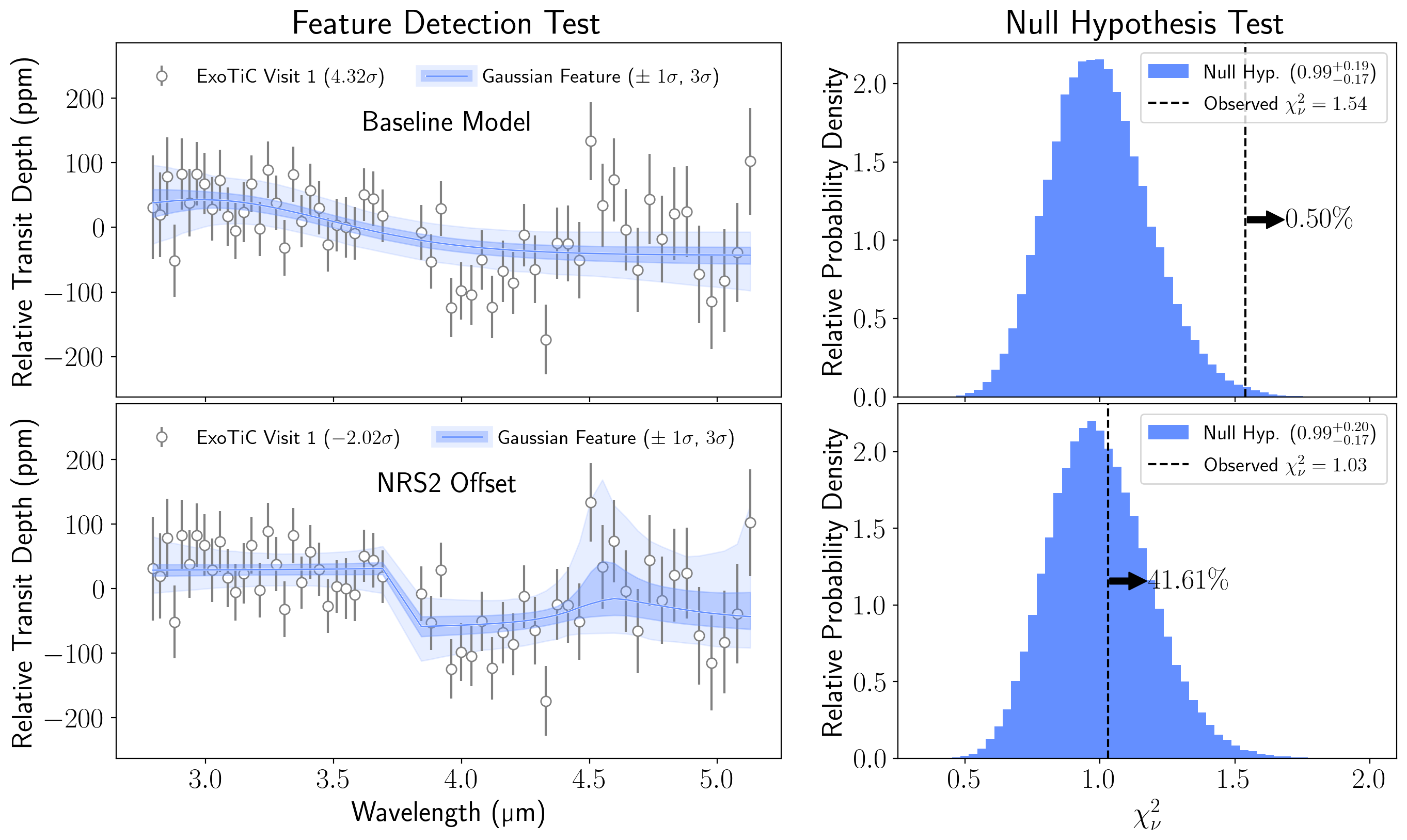}
\figsetgrpnote{\exotic Gaussian Feature Detection Tests for Visit 1 (as shown in Figure~\ref{app:fig:flat}).}
\figsetgrpend

\figsetgrpstart
\figsetgrpnum{1.8}
\figsetgrptitle{\exotic Visit 2}
\figsetplot{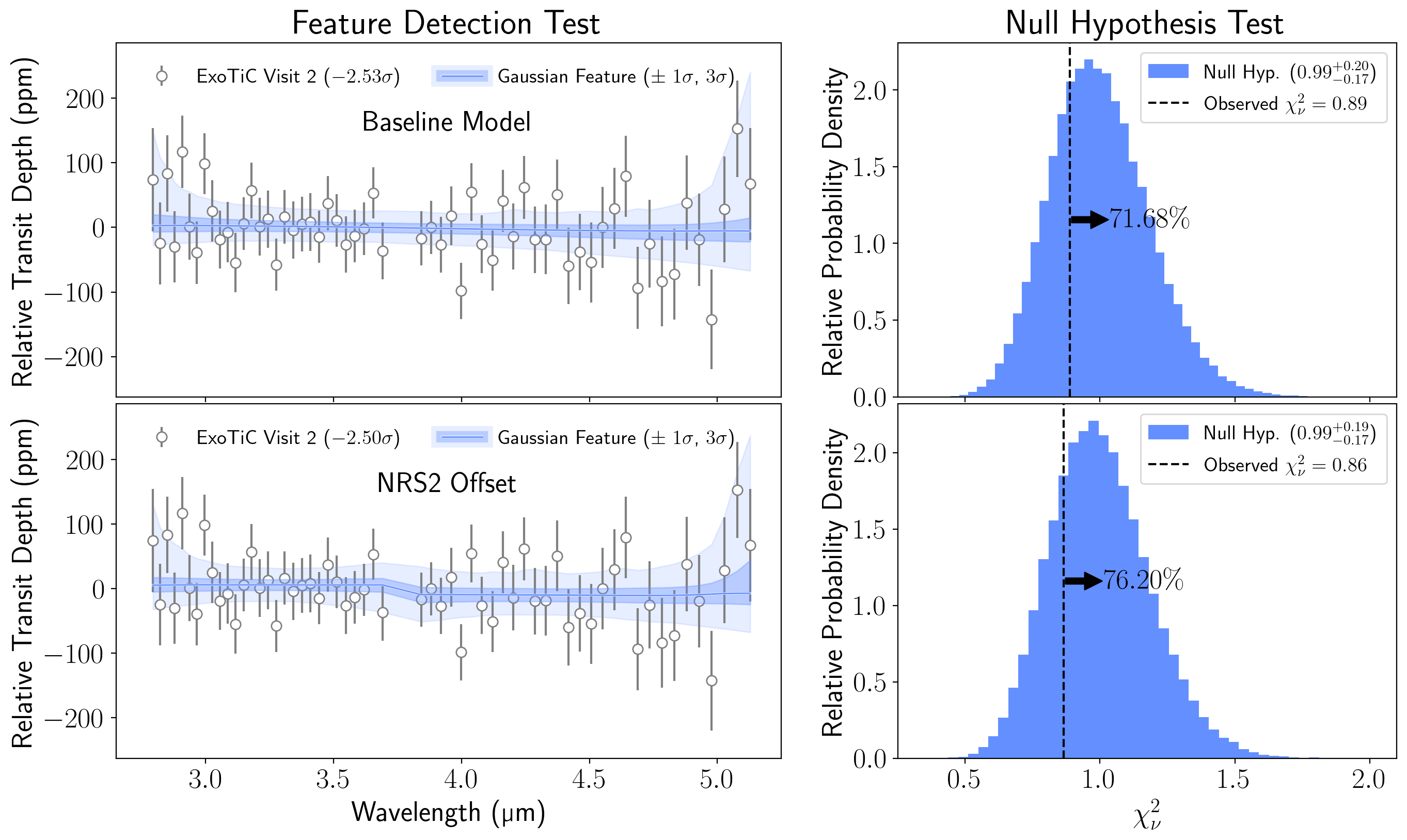}
\figsetgrpnote{\exotic Gaussian Feature Detection Tests for Visit 2 (as shown in Figure~\ref{app:fig:flat}).}
\figsetgrpend

\figsetgrpstart
\figsetgrpnum{1.9}
\figsetgrptitle{\exotic Combined Visit}
\figsetplot{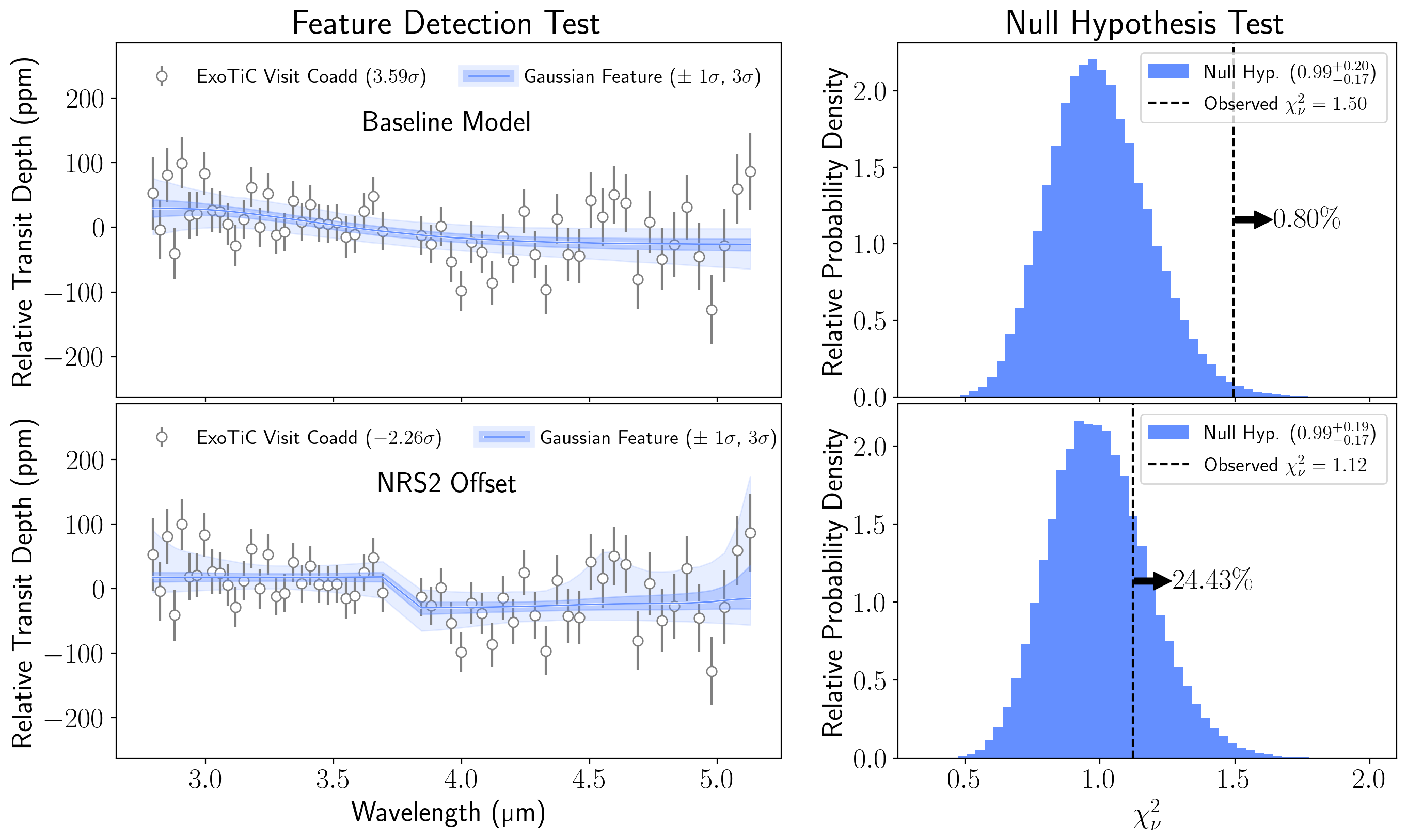}
\figsetgrpnote{\exotic Gaussian Feature Detection Tests for the Combined Visit (as shown in Figure~\ref{app:fig:flat}).}
\figsetgrpend

\figsetend

\pagebreak
\section{Consistency with Previous Data} \label{app:sec:previousdata}

Here we consider the consistency of our JWST NIRSpec G395H transmission spectra of \planetname with previously reported near-infrared and optical observations. We first demonstrate that our JWST data rule out the H$_2$-dominated atmosphere with HCN and aerosols proposed by \cite{Swain2021}. Second, we show that adding the ground-based observations from \cite{DiamondLowe2018} is insufficient to differentiate the atmosphere and starspot scenarios.

\subsection{HST WFC3}

\begin{figure*}[!htb]
    \centering
    \includegraphics[width = \textwidth]
    {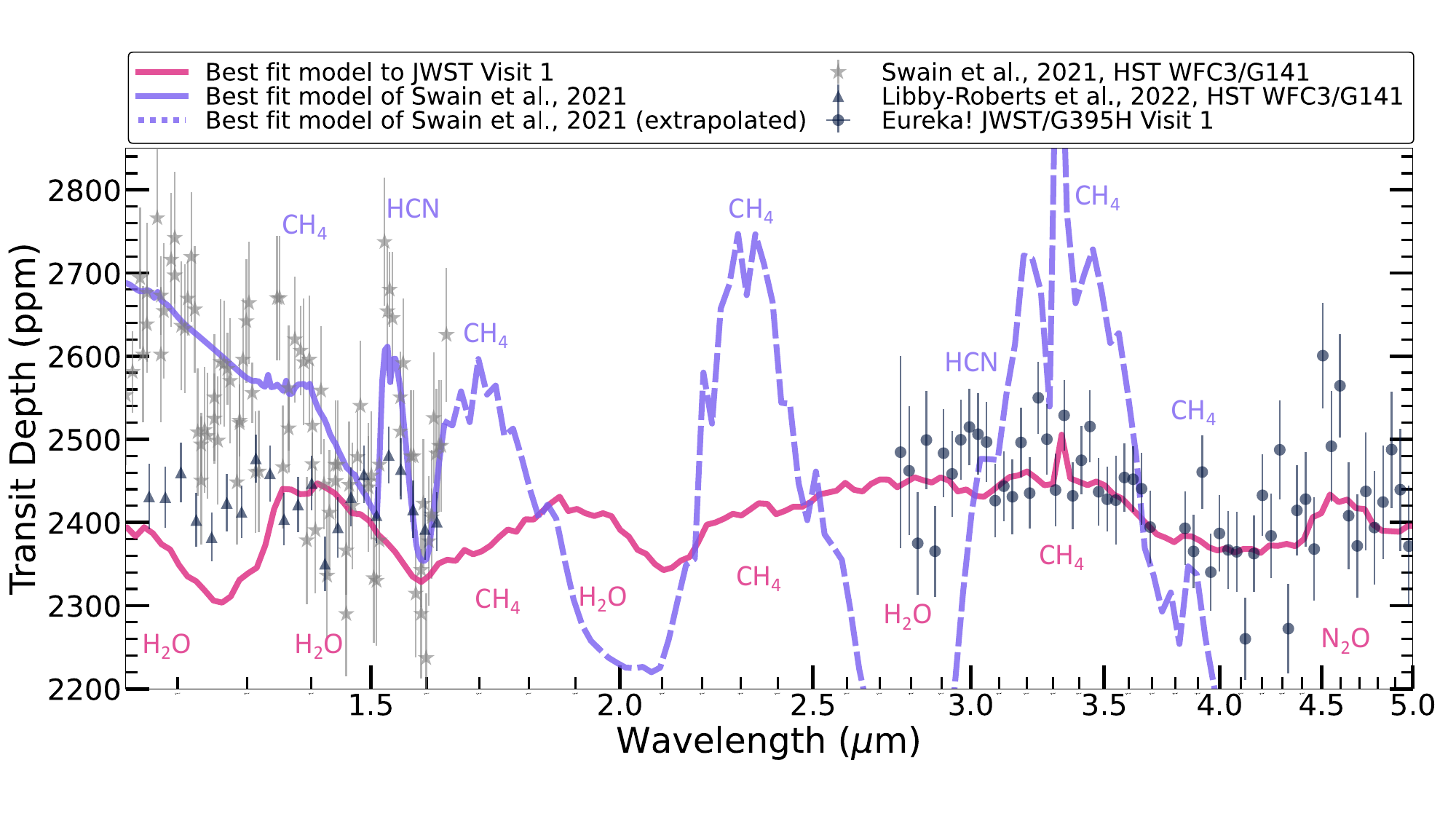}
    \caption{\textbf{Comparison between a proposed H$_2$-dominated spectrum of \planetname and our best-fitting model}. The best-fitting H$_2$O-dominated atmosphere spectrum from the \texttt{POSEIDON} retrieval (pink curve), which we fit to only the \jwst NIRSpec G395H data (black circles), is more consistent with the \hst WFC3 reduction from \citet{LibbyRoberts2022} (black triangles) than the reduction from \citet{Swain2021} (gray stars). The best-fit model from \citet{Swain2021} (purple solid curve), extrapolated out to 5\,$\micron$ with \texttt{PICASO} (purple dashed curve), is inconsistent with our JWST NIRSpec data.}
    \label{app:fig:swain_mod}
\end{figure*}

Figure~\ref{app:fig:swain_mod} compares our best-fitting retrieved spectrum from the atmosphere scenario (pink line, see Section~\ref{ssec:retrievals_atmo}) to both our \jwst data (black circles) and the \hst WFC3 reductions presented in \cite{Swain2021} (gray stars) and \cite{LibbyRoberts2022} (black triangles). Our best-fit model for the \jwst data is inconsistent with the spectral slope and HCN features present by \cite{Swain2021}'s \hst WFC3 reduction. Our best-fitting model is, however, consistent within error (and a modest offset, not applied here) with the reduction presented by \cite{LibbyRoberts2022}. We also show an extrapolated variant (purple dashed line) of the \texttt{Excalibur}\footnote{http://excalibur.ipac.caltech.edu/pages/database} spectral fit from \cite{Swain2021} (purple solid line) extended from 1.7 to 5\,$\micron$ based on their stated model parameters. This model is inconsistent with the atmospheric retrieval of our observed \jwst data, which, as discussed in Section~\ref{ssec:retrievals_atmo}, find no evidence for HCN.

\newpage

\subsection{Ground-Based Data}
    
The ground-based observations of \planetname presented by \cite{DiamondLowe2018} cover important wavelengths (0.7--1.0\,$\micron$) that can potentially differentiate between the atmosphere and starspot scenarios from our \jwst NIRSpec G395H Visit 1 retrievals (see Figure~\ref{fig:retrievals} and Section~\ref{ssec:retrievals_fit_comparison}). However, due to the different telescopes, instruments, data reduction techniques, and the large wavelength gap between the two datasets, one must consider a free offset in retrievals when including both the ground-based and JWST data.

\begin{figure*}[!htb]
    \centering
    \includegraphics[width=\textwidth]{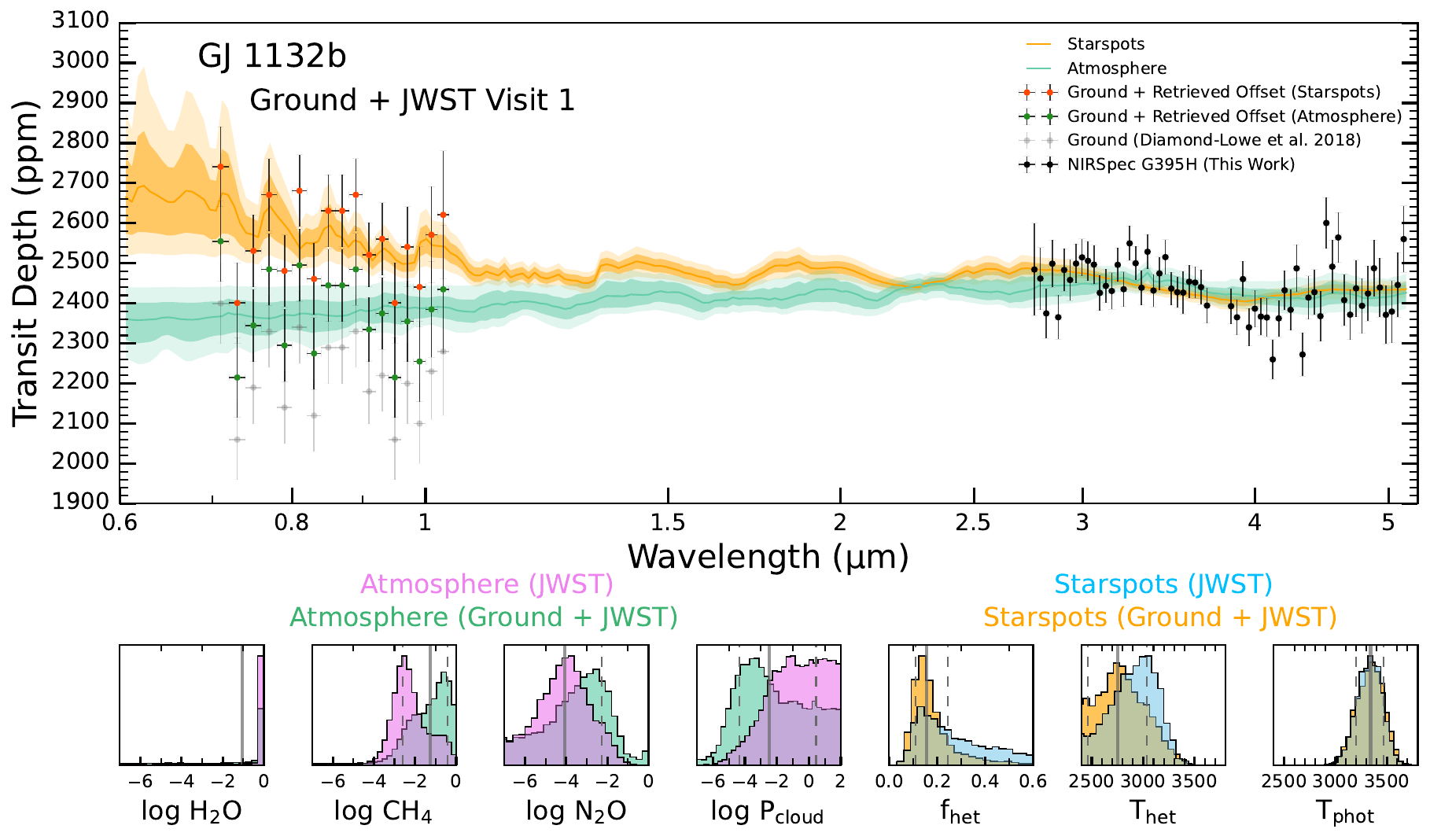}
    \caption{\textbf{Retrieval results including ground-based observations of \planetname}. Top: retrieved transmission spectra for the atmosphere  (Section~\ref{ssec:retrievals_atmo}) and starspot (Section~\ref{ssec:retrievals_spot}) scenarios for JWST NIRSpec G395H Visit 1, but now including the LDSS-3C observations from \citet{DiamondLowe2018}. Both retrievals include a free offset between the ground-based and JWST data. Overplotted are the ground-based observations of \planetname with no offset applied (grey data points), including the median retrieved offset from the starspot scenario (341\,ppm; orange data points), and including the median retrieved offset from the atmosphere scenario (155\,ppm; green data points). Bottom: posterior distributions including the ground-based data (green/orange for the atmosphere/starspot scenarios) compared to the results from Figure~\ref{fig:retrievals} with only JWST data (purple/blue for the atmosphere/starspot scenarios). The retrieval results minimally change when adding the ground-based observations.
    }
    \label{app:fig:retrieval_ground_JWST}
\end{figure*}

Figure~\ref{app:fig:retrieval_ground_JWST} shows our retrieved transmission spectra and atmospheric/stellar properties from two \texttt{POSEIDON} retrievals including the ground-based data from \cite{DiamondLowe2018}. We find that the retrieved properties are broadly consistent with our retrievals only including our JWST NIRSpec G395H data (see Figure~\ref{fig:retrievals}). The only differences are a higher preferred CH$_4$ abundance for the atmosphere scenario and a tighter constraint on the spot coverage fraction for the starspot scenario. However, the inclusion of an offset results allows the ground-based data to move to accommodate either model. Consequently, the ground-based from \cite{DiamondLowe2018} does not allow us to differentiate between the starspot and atmosphere scenarios.

% \pagebreak
% \begin{figure*}
%     \centering
%     \includegraphics[width = \textwidth]{secrets.jpeg}
%     \caption{This Easter egg is only half as mysterious as GJ 1132 b ;) }
%     \label{fig:secrets}
% \end{figure*}

\end{document}